\documentclass[11pt]{article}
\pdfoutput=1
\usepackage{jheparxiv}
\usepackage[latin1]{inputenc}
\usepackage{amsmath}
\usepackage{dsfont}
\usepackage{mathrsfs}
\usepackage{graphicx}
\usepackage{amssymb}
\usepackage{enumerate}
\usepackage[usenames,dvipsnames,svgnames,table,x11names]{xcolor}
\usepackage{cleveref}
\usepackage[colorinlistoftodos]{todonotes}
\usepackage{blkarray}
\usepackage{tikz}
\usepackage{graphicx}

\numberwithin{equation}{section}

\newcommand{\A}{\mathbb{A}}
\newcommand{\C}{\mathbb{C}}

\newcommand{\CP}{\mathbb{CP}}

\newcommand{\PT}{\mathbb{PT}}

\renewcommand{\P}{\mathbb{P}}

\newcommand{\F}{\mathscr{F}}

\newcommand{\scri}{\mathscr{I}}
\newcommand{\M}{\mathbb{M}}

\newcommand{\T}{\mathbb{T}}
\newcommand{\Z}{\mathbb{Z}}
\newcommand{\p}{\partial}
\newcommand{\dbar}{\bar\partial}
\newcommand{\e}{\mathrm{e}}

\newcommand{\cA}{\mathcal{A}}
\newcommand{\cB}{\mathcal{B}}
\newcommand{\cC}{\mathcal{C}}
\newcommand{\cD}{\mathcal{D}}
\newcommand{\cN}{\mathcal{N}}
\newcommand{\cO}{\mathcal{O}}

\newcommand{\cZ}{\mathcal{Z}}
\renewcommand{\P}{\mathbb{P}}

\newcommand{\tr}{\mathrm{tr}}

\newcommand{\rd}{\, \mathrm{d}}

\newcommand{\pf}{\mathrm{Pf}\,}
\newcommand{\be}{\begin{equation}\label}
\newcommand{\ee}{\end{equation}}
\newcommand{\bea}{\begin{eqnarray}\label}
\newcommand{\eea}{\end{eqnarray}}

\newcommand{\la}{\langle}
\newcommand{\ra}{\rangle}

\newcommand{\sA}{{\scalebox{0.6}{$A$}}}
\newcommand{\sB}{{\scalebox{0.6}{$B$}}}
\newcommand{\sC}{{\scalebox{0.6}{$C$}}}
\newcommand{\sD}{{\scalebox{0.6}{$D$}}}
\newcommand{\sI}{{\scalebox{0.6}{$I$}}}
\newcommand{\sJ}{{\scalebox{0.6}{$J$}}}

\title{Ambitwistor Strings in Six and Five Dimensions}

\author[a]{Yvonne Geyer,}
\author[b]{Lionel Mason}
\author[c]{and David Skinner}

\affiliation[a]{Department of Physics, Faculty of Science, Chulalongkorn University\\
Thanon Phayathai, Pathumwan, Bangkok 10330, Thailand}
\emailAdd{yjgeyer@gmail.com}

\affiliation[b]{The Mathematical Institute, University of Oxford\\ AWB, ROQ, Woodstock Rd,  Oxford OX2 6GG, United Kingdom}
\emailAdd{lmason@maths.ox.ac.uk}

\affiliation[c]{Department of Applied Mathematics \& Theoretical Physics, 
University of Cambridge\\
Wilberforce Road, 
Cambridge CB3 0WA, United Kingdom}

\emailAdd{d.b.skinner@damtp.cam.ac.uk}

\abstract{
Ambitwistor strings are chiral (holomorphic) strings whose target is the space of complex null geodesics,  ambitwistor space.  We introduce twistor representations of ambitwistor space in 6 and 5 dimensions.  In 6d the twistor representation is naturally conformally invariant. Anomaly cancellation leads to models that describe biadjoint scalar amplitudes and certain conformally invariant gauge and gravity theories, respectively of $4^{\rm th}$ and $6^{\rm th}$ order. There are three such models, reflecting triality for the conformal group SO(8) associated to these 6d models. On reduction to  five dimensions, gauge anomaly cancellation requires supersymmetry and the resulting models describe maximally supersymmetric Yang-Mills and gravity.  The twistor representation of these ambitwistor strings lead to formul{\ae} for maximally supersymmetric gauge and gravity amplitudes based on the polarized scattering equations in 5d, found earlier by the first two authors.  
}

\begin{document}

\maketitle


\section{Introduction}
Ambitwistor strings \cite{Mason:2013sva} are holomorphic (chiral) string theories  whose target space is  the space $\A$ of complexified\footnote{That is to say, the space-time coordinates are  understood to be analytically continued to complex values, and complex null geodesics have tangent vectors that are complex null vectors.}  null geodesics, known as ambitwistor space. They generalize Witten's twistor string \cite{Witten:2003nn,Berkovits:2004hg} to a variety of theories in different dimensions \cite{Berkovits:2013xba, Geyer:2014fka, Casali:2015vta}.  Their worldsheet correlation functions describe target space scattering amplitudes in a way that is very much analogous to conventional string theories, except that these correlators are supported on the scattering equations, leading, on the Riemann sphere, to tree-level field-theory amplitudes that are rational functions of the external momenta. The original family of ambitwistor strings gave the theory underpinning the CHY formul\ae~\cite{Cachazo:2013iea} describing the scattering of scalars, gluons and gravitons.  The models of \cite{Mason:2013sva} are analogous to the RNS string,  and have an analogue of the Ramond sector that describes fermions (and $p$-form fields in supergravity) \cite{Adamo:2013tsa}. However,  it is not easy to obtain compact formul\ae\ for amplitudes involving these fermions, and target space supersymmetry is not simply expressed. In contrast, the RSVW formul\ae~\cite{Roiban:2004yf}  of the original twistor string incorporated supersymmetry straightforwardly, using twistor geometry to give a completely transparent treatment of fermions and supersymmetry. The purpose of this paper is to present ambitwistor string models six and five dimensions that, by virtue of the use of spinors and twistors, incorporate fermions and supersymmetry in a simple and transparent way. The RNS  ambitwistor string  described ambitwistor space indirectly by gauging the reduction from space-time to the space of null geodesics; fixed vertex operators are built directly from fields on space-time, and the Penrose transform from space-time fields to their representation as  cohomology classes on ambitwistor space is implemented via descent. The presentation we obtain in this paper lives directly in ambitwistor space and vertex operators are built directly from ambitwistor-space cohomology classes.

\begin{figure}[h]
\begin{center}
\begin{tikzpicture} [scale=1.2]
 \fill[Red3, opacity=0.2] (2.364,1.955) -- (2,2.5) -- (-1,1.5) -- (0,0) -- (1,1.5) -- cycle;
  \fill[Red3, opacity=0.2] (0,0) -- (1,-1.5) -- (4,-0.5) -- (3,1) -- cycle;
 \fill[DarkGreen, opacity=0.2] (4,2.5) --  (1,1.5) -- (0,0) -- (3,1) -- cycle;
 \fill[DarkGreen, opacity=0.2] (0,0) -- (-1,-1.5) --(0.636,-0.955) -- cycle;
  \draw[->] (-1.5,-0.5) -- (4.5,1.5);
  \draw[Red3] (2.364,1.955) -- (2,2.5) -- (-1,1.5) -- (0,0) -- (1,-1.5) -- (4,-0.5) -- (3,1);
  \draw[dashed,Red3] (2.364,1.955) -- (3,1);
  \draw[DarkGreen] (3,1) -- (4,2.5) -- (1,1.5) -- (-1,-1.5) -- (0.636,-0.955);
  \draw[dashed,DarkGreen] (0.636,-0.955) -- (2,-0.5) -- (3,1);
  \node at (1.8,-1.6) {\scalebox{0.7}{$\textcolor{Red3}{Z\in {\mathbb{T}}}$}};
    \node at (-0.2,-1.6) {\scalebox{0.7}{$\textcolor{DarkGreen}{W\in {\mathbb{T}^*}}$}};
    \node at (4.7,1.55) {\scalebox{0.7}{$L$}};
 \end{tikzpicture}
 \end{center}
 \caption{Twistor description of ambitwistor space in 4d.}
 \end{figure}
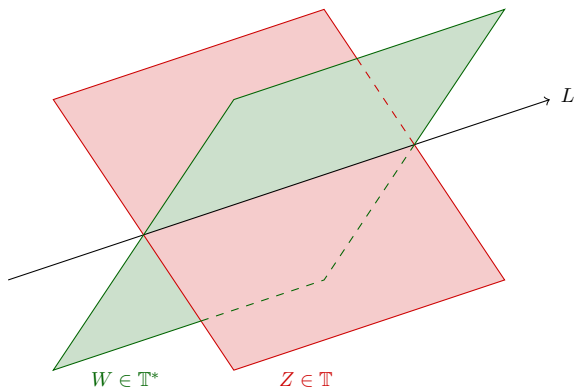

In four dimensions, ambitwistor space $\A$ has a twistorial representation as a reduction of $\T\times \T^*$ where twistor space $\T=\C^{4|\cN}$ can be defined to be the chiral spin space for the superconformal group SL$(4|\cN,\C)$.  More precisely, in four dimensions
$$
\A= \left\{(Z,W)\in \T\times \T^*\ \middle|\  Z\cdot W=0\middle\}\ \middle/\  \C^* \right.
$$
where the $\C^*$ acts as $(Z,W)\sim (\alpha Z, \alpha^{-1} W)$. Berkovits' model \cite{Berkovits:2004hg} for the twistor string has worldsheet fields with values in supertwistors $(Z,W)$ and  action
$$
S=\int_\Sigma W\cdot \bar{D} Z\, .
$$
Here, $\bar{D}$ is a partial connection on a line bundle $\mathcal{L}\to\Sigma$ giving the equivalence $(Z,W)\sim (\alpha Z, \alpha^{-1} W)$. Locally, $\bar{D}=\dbar+a$ and the gauge field $a\in\Omega^{0,1}(\Sigma)$ acts as a Lagrange multiplier imposing the constraint $Z\cdot W=0$. In the quantum theory, this gauging is anomaly free only for the case of maximal $\cN=4$ supersymmetry for super-Yang-Mills and conformal supergravity. A more elaborate version of twistor string theory  that incorporates an analogue of worldsheet supersymmetry has vanishing gauge anomalies for $\cN=8$ and describes Einstein supergravity \cite{Skinner:2013xp}. 

In each of these models, the fields $W$ and $Z$ were originally taken to have different worldsheet conformal weights, slightly obscuring the ambitwistor interpretation. Indeed, when  ${\rm deg}(\mathcal{L})\geq0$ the models localise on maps to twistor space itself, rather than $\A$. Nevertheless, by treating $(Z,W)$ as spinors\footnote{At least at genus zero, this can be interpreted as simply twisting the line bundle $\mathcal{L} \to \mathcal{L}\otimes K^{1/2}_\Sigma$.} on $\Sigma$ an equivalent, but more symmetric description is possible that makes the connection to $\A$ more transparent \cite{Geyer:2014fka}. Furthermore, we will see that it is this ambitwistor interpretation of the twistor string that generalises to higher dimensions: the target space of the theories we construct in this paper are not twistor spaces, but twistor representations of ambitwistor spaces.

There are at least two reasons why the ambitwistor extension is the natural one. Firstly,  in four-dimensions the Penrose transform expresses space-time fields in terms of  cohomology groups $H^{0,1}_{\dbar}$s on twistor space which can easily couple  to a Riemann surface $\Sigma$.  However, in higher dimensions the Penrose transform involves higher degree cohomology classes on twistor space. These higher $(0,p)$-forms do not couple so straightforwardly to a string worldsheet. By contrast, the ambitwistor Penrose transform always allows us to describe space-time fields in terms of $H^{0,1}_{\dbar}$ on $\A$. Secondly, to construct the worldsheet action, the target needs to have a contact structure. The only homogeneous spaces for the conformal groups in higher dimensions that admit contact structures are ambitwistor spaces, see \cite{Boothby:1961} and chapter 2  of \cite{Baston:1989vh}.  

The basic bosonic definition of an ambitwistor as a complex light ray is conformally invariant, and so $\A$ is a homogeneous space for the conformal group. (Ambitwistor strings for field theories that are not conformally invariant arise via the introduction of further worldsheet matter fields that then break conformal invariance.) Twistors have a uniform definition in all dimensions as spinors for the conformal group \cite{Penrose:1986ca}. Therefore it is natural to seek twistor representations of ambitwistor spaces as holomorphic symplectic reductions of linear spaces based on twistors, see for example \cite{Hughston:1988nz,Harnad:1987xq, Harnad:1995zy}.  Ambitwistor strings can then be quantized as the gauged $\beta\gamma$-system realization of this geometry; it was this strategy that led to  \cite{Geyer:2018xgb, Geyer:2019ayz}. Although it was not possible to write down explicit self-contained ambitwistor models for Yang-Mills and gravity in these papers, naive quantization nevertheless led to amplitude formulae for Yang-Mills and gravity.  These are based on an extension of the scattering equations, the polarized scattering equations, that incorporate polarization data. The main obstacle to furnishing complete models in \cite{Geyer:2018xgb, Geyer:2019ayz}  was that the authors were unable to find additional analogues of worldsheet supersymmetry for Yang-Mills and gravity theories in  the chiral framework of these 6d models. 

On reduction to 5d, the chirality issue evaporates and we can identify coherent models. We show here for example that the 5d models do indeed have vanishing worldsheet gauge anomalies as a  consequence of maximal supersymmetry.  We also make some remarks about the original bosonic models in 6d.  These lead to straightforward manifestly conformally invariant  models in 6d.  In 6d, the biadjoint scalar theory with cubic interactions is conformally invariant, and these models compute amplitudes for this theory.  There are also conformally invariant gauge and gravity sectors and we argue that these give rise to the conformally invariant theories identified in \cite{Johansson:2017srf} and give corresponding vertex operators. 

In 6d the complex conformal group is SO$(8,\C)$ and the 6d models have an interesting interplay with triality.  Triality is here expressed in the fact that complexified space-time and the twistor spaces of each chirality are all 6-quadrics, each with inequivalent actions of the conformal group.  Furthermore, ambitwistor space can be expressed as the space of complex projective lines in each of these spaces.  The ambitwistor models can be expressed essentially identically  in each of these spaces, although their vertex operators and coupling to space-time fields are distinct.    The conformally invariant version of the model in the space-time quadric is a twist of that in \cite{Adamo:2017zkm}.

In \S\ref{sec:6d} we introduce the geometry of spinors,  twistors and ambitwistors in six dimensions. In \S\ref{sec:6d-ambitwistor-string} we construct a 6d ambitwistor string based on this geometry. Vertex operators in this model correspond to cohomology classes appearing in the ambitwistor version of the Penrose transform \cite{Baston:1989vh} and are simply related to the twistor representatives for momentum eigenstates constructed in \cite{Mason:2011nw}. These vertex operators describe states in the higher-order gauge and gravitational theories of \cite{Johansson:2017srf} that are conformally invariant in 6d. In the case of the biadjoint scalar, we obtain formul\ae\ for double leading trace amplitudes based on the 6d polarized scattering equations of \cite{Geyer:2018xgb}. These polarized scattering equations were shown to be equivalent to the original biadjoint formulae of CHY \cite{Cachazo:2013hca} in \cite{Albonico:2020mge}. In \S\ref{sec:5d} we symmetry reduce the 6d models down to 5d. By introducing further worldsheet matter we construct twistor versions of ambitwistors strings for 5d Yang-Mills and Einstein gravity.  The worldsheet gauge anomalies vanish in the case of maximal supersymmetry. The genus zero worldsheet correlation functions compute $n$-point tree amplitudes in these theories, with target space supersymmetry manifest. These amplitudes are again given in terms of the polarized scattering equations.

\section{Ambitwistors in six dimensions}
\label{sec:6d}

In this section we provide a brief review of spinors, twistors and ambitwistors in $d=6$, showing that the space of null geodesics can be described via pairs of twistors that obey certain constraints.

\subsection{Spinor-helicity in six dimensions}

In six dimensions, the (complexified) Lorentz group SO$(6,\C)$ is locally isomorphic to SL$(4,\C)$. The  two independent Weyl spinor representations transform in the fundamental and anti-fundamental of SL$(4,\C)$ and we denote them by $\mathbb{S}^\sA$ and $\mathbb{S}_\sA$, where $A=0,\ldots,3$.  The chiral Pauli $\sigma$-matrices are $4\times4$ antisymmetric matrices $\sigma^{\sA\sB}_\mu$ for each $\mu$. This is the highest dimension in which the chiral gamma-matrices provide an isomorphism from Minkowski space vectors to two-index spinors and in 6d these  are skew.  We will therefore  represent six-dimensional vectors as skew spinors  $V^{\sA\sB}  = \gamma^{\sA\sB}_\mu V^\mu$. In particular, complexified space-time $\C^6$ itself may be described with coordinates $x^{\sA\sB} = x^{[\sA\sB]}$ in  which the flat metric is
\begin{equation}
\label{eqn:flat-metric}
\rd s^2 = \frac{1}{2}\,\varepsilon_{\sA\sB\sC\sD}\, \rd x^{\sA\sB}\rd x^{\sC\sD}\,,
\end{equation}
where $\varepsilon_{\sA\sB\sC\sD}=\varepsilon_{[\sA\sB\sC\sD]}$ is the Levi-Civita symbols for SL$(4,\C)$. This metric allows us to raise and lower pairs of skew indices; note that there is no natural way to raise or lower a single spinor index.  

Spinor helicity provides the spinor representation of null momentum vectors and their associated polarization vectors.  A momentum vector $K_{\sA\sB}$ is null  if $\varepsilon^{\sA\sB\sC\sD}K_{\sA\sB}K_{\sC\sD} = 0$. Skew matrices only have even rank, so for $K\neq 0$, it must have rank 2 so that  
$$
K_{\sA\sB} = \frac{1}{2}\epsilon_{ab}\, \kappa_\sA^a \kappa_\sB^b\, , \qquad a=1,2, \qquad \epsilon_{ab}=\epsilon_{[ab]}, \qquad \epsilon_{01}=1\, ,
$$ 
for some pair of spinors $\kappa_\sA^a$,   defined up to SL$(2,\C)$ acting on the $a$ index. We could just as well have written 
$$
K^{\sA\sB}=\frac12 \varepsilon^{\sA\sB\sC\sD}K_{\sC\sD}= \frac{1}{2}\epsilon^{\dot a\dot b}\tilde\kappa^\sA_{\dot a}\tilde\kappa^\sB_{\dot b}, \qquad \dot a=\dot 1,\dot 2\, , \qquad \epsilon_{\dot a\dot b}=\epsilon_{[\dot a\dot b]}\, , \qquad \epsilon_{\dot 0\dot 1}=1\,,
$$ defined up to a distinct copy of SL$(2,\C)$. These $a, \dot a$ indices are those for the little group  SO$(4,\C)\cong {\rm SL}(2,\C)\times{\rm  SL}(2,\C)/\Z_2$ inside the stabilizer of the vector $K^{\sA\sB}$ under the Lorentz group. We often follow 4d spinor-helicity conventions and denote a null $6d$ momentum $K$ as
\begin{equation}\label{little-G}
K_{\sA\sB}={\frac{1}{2}}\epsilon_{ab}\, \kappa_\sA^a \kappa_\sB^b=: {\frac{1}{2}} \langle \kappa_\sA \kappa_\sB\rangle\, , \qquad K^{\sA\sB}= {\frac{1}{2}} \epsilon^{\dot a\dot b}\tilde\kappa^\sA_{\dot a}\tilde\kappa^\sB_{\dot b}=:{\frac{1}{2}} [\tilde\kappa^\sA\, \tilde\kappa^\sB]
\end{equation}  
for some pair of Weyl spinors $\kappa_\sA^a$ defined up to SL$(2,\C)$ transformations. (The factors of $\frac{1}{2}$ are for later convenience.)

\subsection{Twistors in six dimensions}

The twistor space of six-dimensional space-time is defined to be the space of pure spinors for the conformal group. Geometrically,  these correspond to totally null self-dual 3-planes, known as \emph{$\alpha$-planes} ($\beta$-planes being the anti-self-dual ones).  Both chiralities can be represented via pairs of 6d spinors 
\begin{equation}
Z^\cA= (\mu^\sA,\lambda_\sA) \in \mathbb{S}^\sA\oplus \mathbb{S}_\sA\,, \qquad \tilde Z_\cA= (\tilde \mu_\sA,\tilde \lambda^\sA)\in \mathbb{S}_\sA\oplus \mathbb{S}^\sA\, . \label{twistor-ind}
\end{equation}
The conformal group is $SO(8,\C)$ for which  we have triality whereby the chiral and anti-chiral spin representations both have an $SO(8)$ structure, albeit inequivalent to the standard vector one. In particular, there are $SO(8)$-invariant inner products on each chirality of twistor space, defined by 
\begin{equation} \label{eqn:Z-inner-prod}
Z\cdot Z=2\mu^\sA\lambda_\sA \qquad \text{and}\qquad\tilde Z\cdot\tilde Z=2\tilde{\mu}_\sA\tilde\lambda^\sA\,,
\end{equation}
respectively. However,  there is no $SO(8)$-invariant duality between\footnote{The invariant structure that relates $Z$ and $\tilde Z$ is the $SO(8)$ Pauli-matrix that maps $Z\otimes \tilde Z$ to an $SO(8)$ vector.} $Z$ and $\tilde{Z}$.

Our projective twistor space $Q$ will be the projective quadric of null (pure) spinors~\cite{ Penrose:1986ca,Hughston:1988nz,Chern:2009nt,Mason:2011nw, Mason:2012va} defined by
\begin{equation}
	Q = \{\, [Z]\in\CP^7~ | ~ Z\cdot Z = 2\,\mu^\sA\lambda_\sA = 0\,\}\,.
\end{equation}
As in 4d, twistor space is related to space-time via the incidence relations 
\begin{equation}
\label{eqn:incidence}
\mu^\sA = x^{\sA\sB}\lambda_\sB\,.
\end{equation}
These show that a point $x\in\C^6$ corresponds to a $\CP^3\subset Q$ parametrized by $[\lambda_\sA]$. Note that this $\CP^3$ lies inside $Q$, not just $\CP^7$, since $x^{\sA\sB}$ is skew. Conversely, holding $[\mu^\sA,\lambda_\sA]\in Q$ constant, the incidence relation determines a totally null self-dual 3-plane $\alpha\subset \C^6$ by
\begin{equation}
\label{eqn:alpha-plane}
	x_{\sA\sB}(\alpha) = x_{0\sA\sB} +\lambda_{[\sA}\,\alpha_{\sB]}\,,
\end{equation}
parametrized by $\alpha_\sD$ modulo $\lambda_\sD$; these totally null SD 3-planes are the \emph{$\alpha$-planes}. Triality gives such incidence relations also between any pair of space-time,  $\tilde Q$ and $Q$.

To construct worldsheet models for supergravity and super Yang-Mills, we will also be interested in supertwistors (at least after reduction to 5d, discussed below). We let (complexified) chiral  Minkowski superspace $\C^{6|8\mathcal{N}}$ have coordinates $(x^{\sA\sB}, \theta^{\sA \sI})$, $I = 1,\ldots,2\mathcal{N}$. Then if $\CP^{7|2\mathcal{N}}$ has homogeneous coordinates $[\cZ ]= [\mu^\sA,\lambda_\sA,\eta^\sI]$, the  corresponding supertwistor space  is the quadric $Q_{\scalebox{0.6}{$\mathcal{N}$}}$ defined by \cite{Chern:2009nt,Mason:2011nw, Mason:2012va}
\begin{equation}
\cZ\cdot \cZ =2\mu^{\sA}\lambda_\sA +\omega_{IJ}\,\eta^{I}\eta^{J}=0\, .
\end{equation}
Here $\omega_{\sI\sJ}$ is a skew $2\mathcal{N}\times2\mathcal{N}$ matrix that reduces the R-symmetry group to Sp$(2\mathcal{N})$. Extending the bosonic case, supertwistor space is related to chiral Minkowski space via the
incidence relations
\begin{equation}
\label{eqn:super-incidence}
\mu^{\sA}=x^{\sA\sB}\lambda_\sB + \omega_{\sI\sJ}\,\theta^{\sA\sI}\eta^\sJ\, , \qquad\qquad \eta^{\sI}=\theta^{\sA\sI}\lambda_{\sA}\, .
\end{equation}
A fixed point $(x^{\sA\sB}, \theta^{\sA \sI})\in\C^{6|8\cN}$ thus corresponds to a $\CP^3\subset Q_{\scalebox{0.6}{$\cN$}}$ just as in the bosonic case. Conversely, a point $[\cZ]$ in supertwistor space now determines a totally null $3|6\cN$ plane in chiral Minkowski superspace.

\subsection{Ambitwistors in six dimensions}

Generically, two distinct $\alpha$-planes do not intersect. If they are chosen so that they do intersect, they do so in null geodesic.  The $\alpha$-planes corresponding to two twistors $Z^1$ and $Z^2$ intersect iff 
\begin{equation}
\label{eqn:null-line}
	Z^1\cdot Z^2 = \mu^{1\sA}\,\lambda^2_{\sA} + \lambda^1_{\sA}\,\mu^{2\sA} = 0\,.
\end{equation}
This condition implies that the entire line $Z(u) = Z^1 + uZ^2$ lies in the quadric $Q$,  not just in $\CP^7$. When~\eqref{eqn:null-line} holds, the intersection $L = \alpha_1\cap\alpha_2$ of the two $\alpha$-planes is a null geodesic in space-time, parametrized by
\begin{equation}
\label{eqn:null-geodesic}
x_{\sA\sB}(t) = x_{0\sA\sB} + t \,\lambda^1_{[\sA}\,\lambda_{\sB]}^2\,,
\end{equation}
with $\lambda^a_{\sA}$, $a=1,2$ the little group spinor decomposition of the null tangent vector as in \eqref{little-G}. Thus, given a pair $(Z^1,Z^2)\in Q$ obeying $Z^1\cdot Z^2=0$, we can write $\mu^{\sA a}=x^{\sA\sB}\lambda_{\sB}^a$ for some $x^{\sA\sB}$.

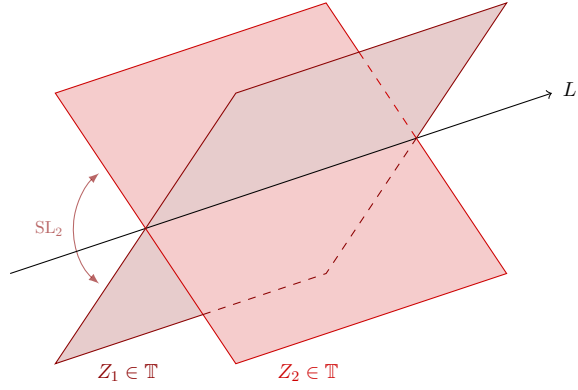
\begin{figure}[h]
\begin{center}
\begin{tikzpicture} [scale=1.2]
 \draw[latex-latex,Red4,opacity=0.6]  (230:0.8) arc(230:130:0.8) node[midway,left]{\scalebox{0.6}{SL$_2$}};
 \fill[Red3, opacity=0.2] (2.364,1.955) -- (2,2.5) -- (-1,1.5) -- (0,0) -- (1,1.5) -- cycle;
  \fill[Red3, opacity=0.2] (0,0) -- (1,-1.5) -- (4,-0.5) -- (3,1) -- cycle;
 \fill[Red4, opacity=0.2] (4,2.5) --  (1,1.5) -- (0,0) -- (3,1) -- cycle;
 \fill[Red4, opacity=0.2] (0,0) -- (-1,-1.5) --(0.636,-0.955) -- cycle;
  \draw[->] (-1.5,-0.5) -- (4.5,1.5);
  \draw[Red3] (2.364,1.955) -- (2,2.5) -- (-1,1.5) -- (0,0) -- (1,-1.5) -- (4,-0.5) -- (3,1);
  \draw[dashed,Red3] (2.364,1.955) -- (3,1);
  \draw[Red4] (3,1) -- (4,2.5) -- (1,1.5) -- (-1,-1.5) -- (0.636,-0.955);
  \draw[dashed,Red4] (0.636,-0.955) -- (2,-0.5) -- (3,1);
  \node at (1.8,-1.6) {\scalebox{0.7}{$\textcolor{Red3}{Z_2\in {\mathbb{T}}}$}};
    \node at (-0.2,-1.6) {\scalebox{0.7}{$\textcolor{Red4}{Z_1\in {\mathbb{T}}}$}};
    \node at (4.7,1.55) {\scalebox{0.7}{$L$}};
 \end{tikzpicture}
 \end{center}
 \caption{Ambitwistor space in 6d parametrized by a pair of twistors $Z_a$.}
 \end{figure}
 
 \medskip

Ambitwistor space $\A$ is the space of such parametrized complex null geodesics, where the parameter comes from the scale of the tangent vector\footnote{Note that since $\lambda_1$ and $\lambda_2$ are part of the homogeneous coordinates for points in $\CP^7$, the parameter $t$ in~\eqref{eqn:null-geodesic} should be taken to scale with weight $(-1,-1)$ under scalings of $(Z_1,Z_2)$.}. We see that in six-dimensions,  $\A$ may be described using a pair of twistors subject to the constraints 
\begin{equation}
\label{eqn:ambitwistor-constraints}
 \A= \left\{ [Z^{a}]\in\C^8\ \middle|\ Z^a\cdot Z^b = 0, \quad  a,b=1,2 \middle\}\ \middle/ \ {\rm SL}(2,\C) \right.
\end{equation}
where the  SL$(2,\C)$ is the little group transformations acting on $a,b$. We have $2\times 8$ variables in the $[Z^a]$, subject to three constraints~\eqref{eqn:ambitwistor-constraints} and quotienting by SL$(2,\C)$  showing that $\A$ is a 10-dimensional complex manifold. This is as expected for the space of parametrized null geodesics: in six dimensions, each light-ray intersects a given five-dimensional Cauchy surface at a unique point, and there is a 5-dimensional family of null vectors through each such point. 

Equation~\eqref{eqn:ambitwistor-constraints} realizes $\A$ as a holomorphic symplectic quotient with respect to the symplectic form  $d\Theta$ on $\T\times \T$ with potential 
\begin{equation}
 \Theta ={\frac{1}{2}} \langle  Z  \cdot dZ\rangle\, .
\end{equation}
It is easily checked that $Z^a\cdot Z^b$ are the Hamiltonians that generate the $SL(2,\C)$ action.  Thus $\Theta$ defines a symplectic potential on $\A$ and is also the reduction of the symplectic potential  $P_{AB}\, dX^{AB}$ from the cotangent bundle of space-time.

It will  be convenient to consider also the space $\P\A$ of \emph{unparametrized} null geodesics, obtained by quotienting $\A$ by an overall scale. Thus, in~\eqref{eqn:null-geodesic}, we consider the null geodesic whose tangent vector is proportional to $\lambda_{[A}^1\lambda^2_{B]}$, considered upto scale.  From above, we see that $\mathbb{P}\A$ is the space of lines in $Q$ and that it inherits a contact structure from the symplectic potential on $\A$.

Ambitwistor spaces with $(0,\cN)$ supersymmetry can be defined completely analogously by replacing $Z^a$ by $\cZ^a$ in the above \cite{Geyer:2018xgb}.  As in the bosonic case, the totally null $3|6\cN$-planes corresponding to two supertwistors $\cZ^1$ and $\cZ^2$ intersect iff $\cZ^1\cdot\cZ^2=0$.  They then intersect along a null super-geodesic $\C^{1|4\cN}$  with tangent vector $(P_{\sA\sB} , \Pi^{\sA\sI})$ satisfying
\begin{equation}
\label{eqn:supermomenta}
	P_{\sA\sB}\propto \langle\lambda_{\sA}\lambda_{\sB}\rangle \,  , \qquad \Pi^{\sA\sI}\lambda_{\sA}^a=0\,.
\end{equation}
Superambitwistor space $\A_{\scalebox{0.7}{$6|2\mathcal{N}$}}$  can thus be represented by pairs of supertwistors $(\cZ^1,\cZ^2)$ that obey $\cZ^a\cdot\cZ^b=0$ and are defined up to little group transformations,
\begin{equation}
 \A_{\scalebox{0.7}{$6|2\mathcal{N}$}} = \left\{ [\cZ^a]\in\CP^{7|2\mathcal{N}}\ \middle|\  \cZ^a\cdot \cZ^b = 0 \middle\}\ \middle/\ {\rm SL}(2,\C)\right. .
\end{equation}
This too is a graded holomorphic symplectic manifold with symplectic potential $\langle \cZ\cdot d\cZ\rangle$.

\medskip

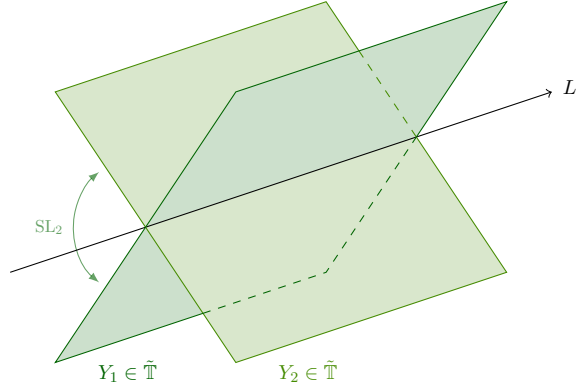
\begin{figure}[h]
 \begin{center}
\begin{tikzpicture} [scale=1.2]
 \draw[latex-latex,DarkGreen,opacity=0.6]  (230:0.8) arc(230:130:0.8) node[midway,left]{\scalebox{0.6}{SL$_2$}};
 \fill[Chartreuse4, opacity=0.2] (2.364,1.955) -- (2,2.5) -- (-1,1.5) -- (0,0) -- (1,1.5) -- cycle;
  \fill[Chartreuse4, opacity=0.2] (0,0) -- (1,-1.5) -- (4,-0.5) -- (3,1) -- cycle;
 \fill[DarkGreen, opacity=0.2] (4,2.5) --  (1,1.5) -- (0,0) -- (3,1) -- cycle;
 \fill[DarkGreen, opacity=0.2] (0,0) -- (-1,-1.5) --(0.636,-0.955) -- cycle;
  \draw[->] (-1.5,-0.5) -- (4.5,1.5);
  \draw[Chartreuse4] (2.364,1.955) -- (2,2.5) -- (-1,1.5) -- (0,0) -- (1,-1.5) -- (4,-0.5) -- (3,1);
  \draw[dashed,Chartreuse4] (2.364,1.955) -- (3,1);
  \draw[DarkGreen] (3,1) -- (4,2.5) -- (1,1.5) -- (-1,-1.5) -- (0.636,-0.955);
  \draw[dashed,DarkGreen] (0.636,-0.955) -- (2,-0.5) -- (3,1);
  \node at (1.8,-1.6) {\scalebox{0.7}{$\textcolor{Chartreuse4}{Y_2\in \tilde{\mathbb{T}}}$}};
    \node at (-0.2,-1.6) {\scalebox{0.7}{$\textcolor{DarkGreen}{Y_1\in \tilde{\mathbb{T}}}$}};
    \node at (4.7,1.55) {\scalebox{0.7}{$L$}};
 \end{tikzpicture}
 \end{center}
 \caption{Ambitwistor space parametrized by a pair of twistors $Y_{\dot a}$ of opposite chirality.}
 \end{figure}

Triality gives three distinct descriptions of $\P\A$ as the space of lines in a 6-quadric.  The second is the  tilded description of $\P\A$ as lines in the opposite chirality twistor space $\tilde Q$.  The third description from triality is the more well-known one arising from the embedding description of conformally compactified Minkowski space $\M$ as a projective quadric $\M\subset \CP^7$.   This arises from extending the space-time coordinates $X=(x^{\sA\sB},s,t)$ with inner product
$$
X\cdot X=x^{\sA\sB}x_{\sA\sB}+2st
$$
and realizing $\M$ as $X\cdot X=0 $ in the projective space. This then gives for ambitwistor space the description\footnote{The orbits of SL$(2,\C)$ acting on $\CP^7$ are  two-dimensional.}
\begin{equation}
\label{ambi-M}
\A= \left\{ [X_m]\in\CP^7\ \middle|\ X_m\cdot X_n = 0, \quad  m,n=1,2 \middle\}\ \middle/ \ {\rm SL}(2,\C)\right..
\end{equation}
This follows because $X_1\cdot X_2=0$ is the condition that points $X_1$ and $X_2 $ are null separated, and the SL$(2,\C)$ quotients the different pairs of points $X_m$ chosen on the null geodesic.

These are all examples of parabolic geometries for the conformal group SO$(8,\C)$.  These are  homogeneous spaces SO$(8,\C)/H$ where $H$ is a parabolic subgroup, i.e., one that contains the Borel subgroup.  Such $H$ are in one-to-one correspondence with subsets of nodes of the Dynkin diagram. The three 6-quadrics related by triality correspond to the outer nodes and $\P\A$ to the central node; see chapter 2 of \cite{Baston:1989vh}.

\section{The worldsheet theory in six dimensions}
\label{sec:6d-ambitwistor-string}

The twistor description of ambitwistor space can be used to construct a new representation of ambitwistor strings that (at least classically) allows us to easily incorporate target space supersymmetry. We begin by showing how a 6d version of the original RNS ambitwistor bosonic string of~\cite{Mason:2013sva} can be reformulated in terms of the geometry of the previous section.

The basic bosonic ambitwistor string action is 
\begin{equation}
S=\int_\Sigma P_\mu \,\dbar X^\mu + \frac{1}{2}\tilde e P^2\,,
\end{equation}
where $\Sigma$ is a Riemann surface and $X:\Sigma\rightarrow M$ is an embedding into complexified space-time. The field $P_\mu$ is a $(1,0)$-form on $\Sigma$, so $P^2$ is a quadratic differential. Therefore, $\tilde e\in \Omega^{0,1}(\Sigma,T_\Sigma)$ transforms as a Beltrami differential and is a Lagrange multiplier enforcing the constraint that $P_\mu$ is null. The corresponding gauge redundancy $X^\mu\sim X^\mu + s P^\mu$ translates $X$ along the null geodesic tangent to $P$. Accounting for the constraint and gauge redundancy, the target space of this model is the space $\P\A$ of null rays, ambitwistor space. Notice that since $P$ takes values in the worldsheet canonical bundle $K_\Sigma$, it is defined only up to scale. 

Specialising to six dimensions, we may solve the constraint $P^2=0$ by setting
\begin{equation}
P_{\sA\sB}={\frac{1}{2}}\epsilon_{ab}\,\lambda_{\sA}^a\lambda_{\sB}^b={\frac{1}{2}}\langle\lambda_\sA\lambda_\sB\rangle
\label{P-lambda}
\end{equation}
for some pair of fields $\lambda_\sA^a$ transforming as (dual) Weyl spinors on space-time. Since $P_{\sA\sB}$ is a section of $K_\Sigma$, these fields are naturally interpreted as sections of $K^{{1}/{2}}_\Sigma$ and so transform as spinors on the worldsheet. With this  $P_{\sA\sB}$, the kinetic term becomes
\begin{equation}
	P_{\sA\sB} \,\dbar X^{\sA\sB}	= {\frac{1}{2}}\epsilon_{ab}\,\lambda^a_\sA\lambda^b_\sB\,\dbar X^{\sA\sB} =
	{\frac{1}{2}}\epsilon_{ab}\left(\lambda_{\sA}^a \,\dbar \mu^{\sA b} +\mu^{\sA b}\,\dbar\lambda^a_\sA\right)
\end{equation}
where we have used the incidence relation~\eqref{eqn:incidence} to introduce a field $\mu^{\sA a} = X^{\sA\sB}\lambda_\sB^a$. Like $\lambda_{\sA}^a$, the fields $\mu^{\sA a}$ also transform as worldsheet spinors. They are naturally invariant under the gauge redundancy $X^{\sA\sB}\mapsto X^{\sA\sB}+s P^{\sA\sB}$.  However, we must now impose the ambitwistor constraints~\eqref{eqn:ambitwistor-constraints} by gauging the little group SL$(2,\C)$ transformations. As discussed after eqn~\eqref{eqn:null-geodesic},  these constraints ensure that $\mu^{\sA a}$ indeed equals $X^{\sA\sB}\lambda^a_\sB$ for some $X^{\sA\sB}$.  We thus write our bosonic ambiwistor string action as
\begin{equation}
\label{eqn:ambitwistor-bosonic-action}
S = {\frac{1}{2}}\int_\Sigma \langle Z\cdot\bar{D}Z\rangle
	= {\frac{1}{2}}\int_\Sigma \epsilon_{ab}\left(\lambda_\sA^a \bar{D} \mu^{\sA b} + \mu^{\sA a}\bar{D}\lambda_\sA^b\right)\,,
\end{equation}
where $\bar{D}Z^b = \dbar Z^b + A^b_{\ c}Z^c$ in terms of a worldsheet $\mathfrak{sl}_2$ gauge field $A^a_{\ b}\in \Omega^{0,1}(\Sigma,\mathfrak{sl}_2)$. (In particular ${\rm tr}(A)=0$, so $A_{ab}=\epsilon_{ac} A^c_{\ b}$ is symmetric in $ab$.) Each $Z^a$ is a section of $K_\Sigma^{1/2}$ and we identify this line bundle with the pullback to $\Sigma$ of the line bundle $\cO(1)\to\CP^{7}$, so that each $Z^a$ indeed describes a map to projective twistor space.  The  $(0,\cN)$ supersymmetric analogues again arise by  replacing the $Z^a$ by fields $\cZ^a$ in the above  \cite{Geyer:2018xgb}. 

We also note that triality gives two further  models, where $Z^a$ is replaced by $\tilde Z_{\dot a}$ defining a line in $\tilde Q$, or by $X_m$, $m=1,2$, defining a line (null geodesic) in $\M$. While these models are formally identical in terms of field content, their relationships to physical space-time are quite distinct. The models in $\tilde Q$ naturally extend to $(\cN,0)$ supersymmetry rather than $(0,\cN)$.  The model in $\M$ is fleshed out further in appendix \ref{M-model}.   This model is a twist of  the 6d conformally invariant model in $\M$ defined by \cite{Adamo:2017zkm}: that model also also gauges three constraints, but uses different choices of worldsheet spin for the pair $X_m$.

\medskip

To promote this theory to a string theory, we allow the complex structure of the worldsheet to vary by replacing $\dbar\mapsto \dbar+e$, where $e\in \Omega^{0,1}(\Sigma,T_\Sigma)$ is a Beltrami differential.  As usual, $e$ couples to the holomorphic worldsheet stress tensor, and we take the path integral to include an integral over the moduli space of inequivalent complex structures (on a possibly punctured Riemann surface).

\subsection{BRST quantization and anomalies} 

As usual, we may handle the gauge redundancies associated to $A_{ab}$ and $e$ by BRST quantization. We introduce fermionic ghosts $n_{ab}=n_{ba}$ and antighosts $m^{ab}$ corresponding to $A_{ab}$, and standard $(b,c)$ ghosts corresponding to $e$. Note that all these fields are chiral on $\Sigma$. At least in the absence of vertex insertions, we can choose the gauge $A_{ab}=0 = e$, in which the gauge-fixed action becomes
\begin{equation}
\label{eqn:ambitwistor-6d}
S  = \int_\Sigma  {\frac{1}{2}}Z_a\cdot\dbar Z^a + m^{ab}\,\dbar n_{ab} + b\,\dbar c\ .
\end{equation}
The corresponding BRST operator is
\begin{equation}
Q = \oint c \left(T+\frac{1}{2}T^{\scalebox{0.6}{$bc$}}\right)+ {\frac{1}{2}}n_{ab}\, Z^a\cdot Z^b + \frac{1}{2} m^{ab} \,\epsilon^{cd}\left[n_{ac},n_{bd}\right]\,,
\end{equation} 
where
\begin{equation}
T= {\frac{1}{2}}Z_a\partial Z^a + m^{ab}\,\partial n_{ab}\,,\qquad T^{\scalebox{0.6}{$bc$}}= (\partial b)c+2b\,\partial c
\end{equation}
is the holomorphic stress tensor.
\medskip

Let us now check that the chiral theory  is non-anomalous. Worldsheet gauge anomalies may be represented by a bubble diagram with two external gauge fields. For SL$(2,\C)$ transformations, including the contribution from the ghosts the coefficient of this diagram is
\begin{equation}
\label{eqn:6d-gauge-anomaly}
a_{\rm SL(2)} = \sum_{i} (-1)^F_i \,{\rm tr}_{R_i}(t^kt^k) = \frac{1}{2}\times 8\,{\rm tr}_F(t^kt^k) - {\rm tr}_{\rm adj}(t^kt^k)\,,
\end{equation}
where the factor of $\frac{1}{2}$ arises since the $Z^a$ are self-conjugate. Using the trace relation ${\rm tr}_{R} (t^kt^k) = C_2(R)\,\dim(R)$ where $C_2$ is the quadratic Casimir, for SL$(2,\C)$ we have
\begin{equation}\label{eq:tr_sl2}
	{\rm tr}_F(t^kt^k) = \frac{3}{2}\qquad\qquad{\rm tr}_{\rm adj}(t^kt^k)= 6\,,
\end{equation}
so the SL$(2,\C)$ anomaly~\eqref{eqn:6d-gauge-anomaly} vanishes. 

We remark that the anomaly counting in the models related by triality is identical,  and also in the model of \cite{Adamo:2017zkm} related to these by a twist. We also remark that, without the incorporation of further fields, SL$(2,\C)$ anomaly cancellation rules out the natural supersymmetry that arises by replacing $Z^a$ with $\cZ^a$.  The 5d ambitwistor strings we consider below involve additional fields which indeed modify the condition for anomaly cancellation.

At this stage, the central charge of our theory is
\begin{equation}
	{\rm c} = -8_Z - 6_{mn}-26_{bc}\,
\end{equation}
so for the Virasoro anomaly to vanish, we must couple~\eqref{eqn:ambitwistor-6d} to a further internal CFT of central charge ${\rm c}=40$. In the following section, we will do this by including (two) worldsheet current algebras.

\subsection{Vertex operators and the 6D polarized scattering equations}\label{sec:6d_VO}

We briefly review the discussion in \cite{Geyer:2018xgb, Albonico:2020mge}.  Massless fields on space-time can be represented by Dolbeault cohomology classes $H^1_{\dbar}(\P\A, E)$ via the ambitwistor Penrose transform, where the bundle $E$ depends on the type of field. Unlike the Penrose transform for twistor space, here the correspondence works off-shell. In the RNS model, the field equation instead arises from requiring that vertex operator built from this cohomology class is BRST closed at the quantum level. In our twistor description of ambitwistor strings, we will instead construct vertex operators starting from the representative of a plane wave on twistor space $Q$, which is necessarily on-shell.

It has been known for many years that solutions to massless field equations in 6d Minkowski space can be obtained from cohomology classes of degree 2 or 3 on $Q$ \cite{Baston:1989vh,Chern:2009nt,Mason:2011nw, Mason:2012va}.  In \cite{Mason:2011nw} in particular, representatives in $H^2$ and $H^3$ were constructed for momentum eigenstates with polarization data given in the 6d spinor-helicity format.  The simplest example is 
\begin{equation}\label{twistor-rep}
\phi_k(Z)=\int d^2v \, \bar\delta^4(\lambda_A-\langle v\kappa _A\rangle)\, \exp\left(\frac{ \mu^\sA\la\kappa_{\sA}\epsilon\ra}{\langle \epsilon v\rangle}\right) \ \in \ H^2(Q,\cO(-2))\,,
\end{equation}
corresponding to an on-shell massless scalar of momentum $k_{\sA\sB}=\frac{1}{2}\la\kappa_\sA\kappa_\sB\ra$. Here, the holomorphic delta functions are defined by
\begin{equation}
\bar\delta(z)=\frac{1}{2\pi i} \dbar \frac 1z = \delta(\Re z)\,\delta(\Im z) \,d\bar z\,.
\end{equation}
Two of these $\bar\delta$-functions can be used to perform the integrals over $v^a\in\C^2$, with the remaining two giving $\phi_k(Z)$ its cohomology degree. We have also introduced an arbitrary $\epsilon^a\in\C^2$ in the exponential. The idea here, as in the twistor space of 4d space-time, is that on the support of the $\bar\delta$-functions,  the exponential becomes the standard plane wave factor $\e^{ik\cdot x}$ when we use the
twistor incidence relation $\mu^\sA = x^{\sA\sB}\lambda_\sB$. 

To build a plane wave representative on ambitwistor space, we Penrose transform~\eqref{twistor-rep} to $\P\A$ by integrating it over the $\CP^1\subset Q$ corresponding to a point in $\P\A$. Letting $u_a$ be homogeneous coordinates along this $\CP^1$, on $\P\A$ we have 
\begin{equation}\label{mom-estate}
\begin{aligned}
\Phi_k(Z^a) &= \int \la u\,du\ra \,\phi_k(Z(u))
	=\int \la u\,du\ra\,d^2v \,\bar\delta^4(\la u\lambda_\sA\ra - \la v\kappa_\sA\ra) \,\exp\left(\frac{\la u\mu^\sA\ra\la \kappa_\sA\epsilon\ra} {\la\epsilon v\ra}\right)\\
&=\int d^2u\, d^2v\, \bar\delta\big(\langle v \epsilon \rangle -1 \big)\,\bar \delta^4\Big(\langle u\lambda_A\rangle -\langle v\kappa_{A}\rangle\Big)\,\e^{ i\la u\mu^{A}\ra \epsilon_{A}}\, ,
\end{aligned}
\end{equation}
where 
\begin{equation}\label{PA-PT-inc}
	Z(u)=Z_au^a=\langle Z u\rangle
\end{equation}
is the incidence relation between $Q$ and $\P\A$ and $\epsilon_\sA = \la\kappa_{\sA}\epsilon\ra$.  (For fields of spin $s>0$, we will see later that $\epsilon_\sA $ can be identified with the polarization data.) In going to the second line of~\eqref{mom-estate}, we have included an extra integral over the scale of $u$. The  $\bar\delta$-functions relate this scale to that of $v$, which is fixed by the additional factor of $\bar\delta(\la v\epsilon\ra-1)$. 

The integrals over the auxiliary parameters $(u,v)$ can be done against four of the $\bar\delta$-functions, with one $\bar\delta$-function remaining. This shows that $\Phi_k\in H^1(\P\A,\cO(-2))$, which is the correct cohomology group to describe scalar fields on $\P\A$; the twistor construction ensures this field is on-shell. It was shown in \S3 of \cite{Albonico:2020mge} that $\Phi_k$ corresponds to the factor $\bar\delta(k\cdot P)  \,\e^{ik\cdot x}$ of the RNS ambitwistor string.  In particular, we will see why  the remaining $\bar\delta$-function is a multiple of $\bar\delta(k\cdot P)$ in the discussion following \eqref{polscatt}. Notice that the integral over $u$ ensures that $\Phi_k$ is invariant under SL$(2,\C)$ transformations of the $Z^a$. Also note that the OPE of $\la Zu\ra$ with itself is regular, so $\Phi_k$ is free of normal ordering ambiguities.

\medskip

Since the fields $Z^a(\sigma)$ are worldsheet spinors, when pulled back to the worldsheet, $\Phi_k(\sigma)\in H^1(\Sigma,T_\Sigma)$. Thus we can use $\Phi_k$ to construct a (1,1)-form vertex operator $V$ as
\begin{equation}\label{int-vertex}
	V =  w(\sigma)\, \Phi_k(\sigma)
\end{equation}
for some $w\in\Omega^0(\Sigma,K^2_{\scalebox{0.6}{$\Sigma$}})$. Depending on the target space theory, this $w$ may be built from the basic ambitwistor fields $Z$, or could involve additional worldsheet matter. The simplest choice for $w$ is
\begin{equation}\label{eqn:scalar-Vop}
w_{\scalebox{0.7}{scalar}}=t_{\mathfrak{a}}  j^{\mathfrak{a}}\,\tilde t _{\dot{\mathfrak{a}}} \tilde \jmath ^{\dot{\mathfrak{a}}}\,,
\end{equation}
describing a bi-adjoint scalar. Here, $j^{\mathfrak{a}}$ and  $\tilde \jmath ^{\dot{\mathfrak{a}}}$ define two current algebras on the worldsheet (independent of $\P\A$), while $(t_{\mathfrak{a}},\tilde t_{\dot{\mathfrak{a}}})$ label the colours of the external scalar. For an anomaly-free theory these current algebras should provide central charge ${\rm c}=40$.  The double-leading trace part of the correlator of $n$ such states generates tree amplitudes in the cubic bi-adjoint scalar theory; in 6d this theory is conformally invariant at the classical level.

The theory also contains vertex operators describing gluons and gravitons, whose polarization structure can be built from the basic ambitwistor field $Z^{\cA a}$, $\cA=1,\ldots , 8$ as in \eqref{twistor-ind} and $a=1,2$ as in \eqref{eqn:ambitwistor-constraints}. These are
\begin{equation}
w_{\scalebox{0.7}{gauge}}=  E_{\cA\cB} \langle Z^\cA Z^{\cB}\rangle \,t_{\mathfrak{a}}j^{\mathfrak{a}} \, , \qquad 
w_{\scalebox{0.7}{grav}}= E_{\cA\cB} \langle Z^\cA Z^{\cB}\rangle \,\tilde E_{\cC\cD} \langle Z^\cC Z^{\cD}\rangle\, , \label{eqn:DF2-Vop}
\end{equation}
where the $E_{\cA\cB}=E_{[\cA\cB]}$ label the polarization. To ensure these states are BRST closed (and free of normal ordering ambiguities), the polarization structures $E_{\cA\cB}$ and $\tilde E_{\cA\cB}$ must obey a transversality constraint. If $E_{\cA\cB}$  is chosen to project onto the $\lambda^a_A$ spinor part of $Z^\cA$, the polarization structure becomes
\begin{equation}\label{ordinarycase}
E_{\cA\cB} \langle Z^\cA Z^{\cB}\rangle = \frac{1}{2}\epsilon^{\sA\sB}\langle \lambda_\sA \lambda_{\sB}\rangle = \epsilon^{\sA\sB}P_{\sA\sB}
\end{equation}
so that the vertex operators~\eqref{eqn:DF2-Vop} reduce to those of the original bosonic ambitwistor string of \cite{Mason:2013sva}. In this case, the transversality condition is the standard $\epsilon^{AB}k_{AB}=0$.

More generally, note that $J^{\cA\cB} = \langle Z^\cA Z^{\cB}\rangle$ furnish a current algebra for the conformal group SO(8). The general transversality constraint is\footnote{Here the twistor index on the derivative has been raised by the twistor inner product used in the worldsheet kinetic term.}
\begin{equation}\label{eq:conf_cons_rels}
E_{\cA\cB} \langle Z^\cA\, \p^\cB\Phi_k\ra=0\, .
\end{equation}
so that the element of $\mathfrak{so}_8(\C)$ selected by the polarization tensor $E_{\cA\cB}$ preserves the momentum eigenstate. The case \eqref{ordinarycase} corresponds to a translation,  but any conformal generator that preserves $k_{\sA\sB}$ (such as a rotation around $k_{\sA\sB}$) will also give a consistent vertex operator.

We can understand these additional states if we recognise that (e.g.)  the vertex operator  $w_{\scalebox{0.7}{gauge}}\Phi_k$ is obtained from the twistor representative
\begin{equation}
	\phi_k(Z)\, E_{\cA\cB} Z^\cA dZ^\cB\; \in H^2(Q,\Omega^1)
\end{equation}
simply by integration of this $(2,1)$-form over the sphere $Z(u)=\langle Z u\rangle$. More generally, any  $A\in H^2(Q,\Omega^1)$ can be used to generate an element of $H^1(\P\A)$ by integration over the corresponding $\CP^1$ given by \eqref{PA-PT-inc}, $\int_{\CP^1} A$.  Furthermore, if $A= d \gamma$ where $\gamma\in H^2(Q,\cO)$, then the integral will vanish by Stokes theorem.  Thus, by coupling them to one copy of the internal current algebra,  we can construct gauge theory vertex operators from $H^2(Q, \Omega^1/d \Omega^0)$. Writing $A= A_\cA dZ^\cA$, the consistency of the vertex operators constructed in this way  requires $\p^\cA A_\cA=0$. However, $A$ is defined only modulo the addition of $f(Z) Z\cdot dZ$ which vanishes on the quadric $Q=\{Z\cdot Z=0\}$, and we can always choose $f$ to ensure $\p^\cA A_\cA$ vanishes.

\medskip

Following the work of \cite{Azevedo:2017lkz,Berkovits:2018jvm,Azevedo:2019zbn}, the ambitwistor string correlator of states $w_{\scalebox{0.6}{gauge}}\Phi_k$ with standard polarization structure~\eqref{ordinarycase} are understood to correspond to amplitudes in the 4$^{\rm th}$-order gauge theory with action 
\begin{equation}\label{4th-gauge}
	\int {\rm tr}(D^\dagger F\wedge \star\, D^\dagger F)
\end{equation}
on space-time. Using these polarizations in $w_{\scalebox{0.6}{grav}}$ similarly corresponds its gravitational double copy which is a 6$^{\rm th}$-order theory \cite{Johansson:2017srf}. In six dimensions, both these higher derivative theories are conformally invariant. The vertex operators in~\eqref{eqn:DF2-Vop} for general polarizations $E_{\cA\cB}$ give the conformally invariant completion.  It is reasonable to conjecture that $H^2(Q, \Omega^1/d \Omega^0)$ is mapped by the twistor Penrose transform to solutions of the $4^{\rm th}$-order conformally  invariant Maxwell operator $G_2$ given in equation (6) of Branson-Gover \cite{Branson:2004}, and that this operator corresponds to the linearized equations $\Box \,d^\dagger F=0$ of the Johansson-Nohle theory~\eqref{4th-gauge} that is conformally invariant classically in 6d. (The Branson-Gover operator $G_2$ and the Penrose transform are also both conformally invariant.) The corresponding space-time gauge fields are $A_\mu(x)=V_\mu(x)\,\e^{ik\cdot x}$ where $V$ is a conformal Killing vector satisfying  $V(k\cdot x)=0$. This $4^{\rm th}$-order theory thus contains states beyond those of Maxwell's equations.  The 6$^{\rm th}$-order gravitational theory likewise contains linearized states $h_{\mu\nu(x)} = V_{(\mu}W_{\nu)}\,\e^{ik\cdot x}$ for any pair of such conformal Killing vector fields $(V,W)$.

\medskip

It is worth pointing out that \cite{Geyer:2018xgb,Albonico:2020mge}  obtained integrand  formul\ae\  for many other theories in 6d, including supersymmetric gauge and gravity theories. At present the worldsheet theories required to produce these formul\ae\ remains elusive in 6d.

\section{Ambitwistor strings in five dimensions}
\label{sec:5d}

In this section we develop a 5d ambitwistor string by reducing the 6d models of \cref{sec:6d}, gauging a translation symmetry to reduce the target space to the ambitwistor space for 5d Minkowski space.  The worldsheet matter choices for the 6d models naturally lead to models for the bi-adjoint scalar and the reduced conformal Yang-Mills theory. 
Using the supersymmetric extension of ambitwistor space (and including some additional worldsheet matter) we then construct 5d ambitwistor strings that describe maximally supersymmetric Einstein gravity and maximally supersymmetric Yang-Mills theory in five dimensions.  The worldsheet correlators of these models are shown to reproduce the  manifestly supersymmetric formul\ae\ for 
5d tree-level amplitudes from \cite{Albonico:2020mge}.

\subsection{Spinors, twistors and ambitwistors in five dimensions}
\label{sec:5d-reduction}

Spinors in five dimensions reduce directly from the 6d spinors reviewed in the last section. Choose a fixed, non-null 6-vector $\Omega^{\mu}$, with spinor form $\Omega^{\sA\sB}=\Omega^{[\sA\sB]}$, pointing in the `extra' 6th dimension. We then take our complex 5d Minkowski space $\C^5$ to be a plane in $\C^6$ that is orthogonal to $\Omega^{\sA\sB}$, so that tangent vectors $V^{\sA\sB}$ in five dimensions obey $\epsilon_{\sA\sB\sC\sD}\,\Omega^{\sA\sB}V^{\sC\sD}=0$. Fixing the skew form $\Omega^{\sA\sB}$ breaks the $6d$ spin group SL$(4,\C)$ to Sp$(4,\C)\cong{\rm Spin}(5,\C)$. In contradistinction to 6d,  we can now use $\Omega^{\sA\sB}$ and $\Omega_{\sA\sB} = \frac{1}{2}\epsilon_{\sA\sB\sC\sD}\,\Omega^{\sC\sD}$ to raise and lower individual spinor indices, reflecting the lack of chiral spinors in odd dimensions. 

While  a 5d null momentum $k_{\sA\sB}$ may still be described as $k_{\sA\sB} = \frac{1}{2}\langle\kappa_\sA\kappa_\sB\rangle$ in terms of a pair of spinors $\kappa_\sA^a$, the $\kappa_\sA^a$ are now constrained by 
\begin{equation}
\label{eqn:momenta}
\Omega\cdot k = \frac{1}{2}\Omega^{\sA\sB}\epsilon_{ab}\,\kappa_\sA^a\kappa_\sB^b =
 \frac{1}{2}\langle\kappa^\sA\kappa_\sA\rangle  =0\,.
\end{equation}
This condition gives a co-dimension-1 subspace of degree 2 inside the ambitwistor space of six dimensions.  To obtain the ambitwistor space for 5d, the space of null geodesics in (flat) five dimensional space-time, we must also quotient by translations along the extra dimension: the condition $\langle\lambda^\sA\lambda_\sA\rangle=0$ ensures that the tangent vector $p_{\sA\sB} = \langle \lambda_\sA\lambda_\sB\rangle$ to any null geodesic is parallel to the chosen 5-plane, but we must also  identify parallel null geodesics in 6d that lie in different such parallel 5-planes.  We therefore  quotient by the translations along $\Omega^{\sA\sB}\,\p/\p x^{\sA\sB}$ that translate the orthogonal 5-planes into one another. This is simply the Hamiltonian vector field corresponding to $\Omega\cdot p$ via the symplectic form $\rd p_{\sA\sB}\wedge\rd x^{\sA\sB}$.

Equivalently, we can implement the reduction on the pair of twistors  $[Z^a]$, using the definition \eqref{eqn:ambitwistor-constraints} of ambitwistor space in 6d. Using the symplectic form $\epsilon_{ab}\,\rd \mu^{Aa}\wedge\rd \lambda_A^b$ instead of $\rd p_{\sA\sB}\wedge\rd x^{\sA\sB}$, the constraint 
$\langle\lambda^\sA \lambda_\sA\rangle=0$    generates the transformations 
\begin{equation}\label{eq:transf_5d}
	\mu^{\sA a} \mapsto \mu^{\sA a} + r \,\Omega^{\sA\sB}\lambda_\sB^a\,,\qquad \qquad \lambda^a_\sA\mapsto \lambda^a_\sA\,.
\end{equation}
These are generated by the Hamiltonian vector field $\Omega^{\sA\sB}\lambda_\sA^a\,\p/\p\mu^{\sB a}$ with Hamiltonian $\langle\lambda^\sA\lambda_\sA\rangle$;  that this corresponds to the translations along the sixth dimension $\Omega^{\sA\sB}\,\p/\p x^{\sA\sB}$ can be verified directly from the incidence relations \eqref{eqn:incidence}. Note that the transformations \eqref{eq:transf_5d} preserve the constraints $Z^a\cdot Z^b=0$ since $\Omega^{\sA\sB}$ is skew and so are defined on $\mathbb{A}_6$. Thus we have
\begin{equation*}
 \A_5 = \left\{ Z^a\in\T\times \T \  \middle|\ Z^a\cdot Z^b = 0, \  \langle\lambda^\sA\lambda_\sA\rangle=0 \middle\}\middle/ \middle\{\mathrm{SL}(2,\C)\times \C\right\}\,,
\end{equation*}
where the $\C$-factor in the quotient is the action of the translation \eqref{eq:transf_5d}.
Accounting for this additional constraint and redundancy, we see that the ambitwistor space $\A_5$ of five dimensions is an eight-dimensional complex manifold.  As before this reduction is a holomorphic symplectic quotient with respect to the symplectic form $d\Theta$ generated by the Hamiltonians $Z^a\cdot Z^b$ and $\langle \lambda^A\lambda_A\rangle$.

The reduction to the 5d superambitwistor space $\A_{\scalebox{0.7}{$5|2\mathcal{N}$}}$ is now straightforward, using the same symplectic quotient by  $\langle\lambda^\sA\lambda_\sA\rangle=0$ as in the bosonic case and replacing $Z^a$ by $\cZ^a$.

\subsection{Worldsheet ambitwistor theories for five dimensions}
\label{sec:5d-ambitwistor-string}

The symplectic reduction of ambitwistor space from 6d to 5d is realized in the ambitwistor string by an additional gauging of the 6d bosonic action \eqref{eqn:ambitwistor-bosonic-action}.  The current $\langle\lambda^\sA\lambda_\sA\rangle$ is the pullback to $\Sigma$ of the Hamiltonian generating translations along $\Omega^{\sA\sB}\p/\p x^{\sA\sB}$. As usual, the gauge field  $a\in\Omega^{0,1}(\Sigma)$ acts  as a Lagrange multiplier imposing  the constraint $\langle\lambda^\sA\lambda_\sA\rangle=0$, while the associated gauge transformations identify the hypersurfaces orthogonal to $\Omega^{\sA\sB}$.  The bosonic 5d ambitwistor action is thus
\begin{equation}\label{eq:S_bos_5d}
S_{\scalebox{0.6}{$5d$}}^{\scalebox{0.6}{bos}}=\int_\Sigma {\frac{1}{2}}Z\cdot\bar{D} Z + a\, \langle\lambda^\sA\lambda_\sA\rangle
= \int_\Sigma {\frac{1}{2}}\epsilon_{ab}\left(\lambda^a_\sA\bar {D}\mu^{\sA b} + \mu^{\sA a}\bar{D}\lambda_\sA^b\right) + a \langle\lambda^\sA\lambda_\sA\rangle\,.
\end{equation}
As in 6d, we have used the notation $\bar{D}Z^a  =(\dbar+e\partial) Z^a + A^a_{\ b}Z^b$, where $A^a_{\ b}\in\Omega^{0,1}(\Sigma,\mathfrak{sl}_2)$ is a worldsheet $\mathfrak{sl}_2$ gauge field and $e \in\Omega^{0,1}(\Sigma,T_\Sigma)$ is a Beltrami differential gauging deformations of the worldsheet complex structure.

Just as in 6d, we can construct a critical model by coupling this bosonic ambitwistor action to an internal CFT, e.g. in the form of a current algebra. The resulting model will contain bi-adjoint scalar states, as well as (the dimensional reduction of) the higher order gauge and gravity states discussed above. Such a theory corresponds to the dimensional reduction of the theories of \cite{Johansson:2017srf}.

However, in 5d we also have the more interesting possibility to construct worldsheet models for conventional maximally supersymmetric Yang-Mills theory and Einstein gravity. 
Both models are based on the natural supersymmetric extension of the bosonic action \eqref{eq:S_bos_5d},
\begin{equation}\label{eqn:5d-supertwistor-action}
 S_{\scalebox{0.6}{$5d$}}^{\scalebox{0.6}{$\mathcal{N}$}}=\int_\Sigma {\frac{1}{2}}\cZ\cdot\bar{D} \cZ + a\,\langle\lambda^\sA\lambda_\sA\rangle\,.
\end{equation}
Here, the supertwistor fields $ \cZ^a$ are again worldsheet spinors,
and the Lagrange multiplier term  $a\,\langle\lambda^\sA\lambda_\sA\rangle$ implements the symplectic quotient to $\P\A_{\scalebox{0.7}{$5|2\mathcal{N}$}}$ as discussed in \S\ref{sec:5d-reduction}. The action $S_{\scalebox{0.6}{$5d$}}^{\scalebox{0.6}{$\mathcal{N}$}}$ agrees with the familiar Green-Schwarz ambitwistor string action reduced to five dimensions,
\begin{equation}
S=\int_\Sigma P_{\sA\sB}\left(\dbar X^{\sA\sB}+ \frac{1}{2} \theta^\sA_\sI\,\dbar \theta^{\sB\sI}\right) + \frac{1}{2}\tilde e\, P^2 + a\, \Omega^{\sA\sB}P_{\sA\sB}\,,
\end{equation}
using $P_{\sA\sB} = \epsilon_{ab}\,\lambda_\sA^a\lambda_\sB^b/2$ as well as  the super incidence relations~\eqref{eqn:super-incidence}.

To obtain  ambitwistor strings describing 5d super Yang-Mills or Einstein supergravity, we must couple~\eqref{eqn:5d-supertwistor-action}  to further fields. As proposed in \cite{Geyer:2018xgb}, we introduce a pair of worldsheet fermions $\rho_\sA,\tilde\rho_\sA \in\Pi\Omega^0(\Sigma,K^{1/2}_\Sigma\otimes \mathbb{S}_\sA)$, with action\footnote{We also discuss an equivalent action in \cref{app:rho}, with both fermions combined into one field $\rho_\sA^a$ with an additional SL$(2,\mathbb{C})$ index.} 
\begin{equation}
\label{eqn:rho-action}
S_{\rho}= \int_\Sigma \tilde\rho^\sA\,\dbar\rho_\sA +\chi_a\,\lambda^{a}_\sA\rho^\sA+  \tilde \chi_a\,\lambda^a_\sA\tilde \rho^\sA\,,
\end{equation}
where $\tilde\rho^\sA=\Omega^{\sA\sB}\tilde\rho_\sB$ and $\rho^\sA=\Omega^{\sA\sB}\rho_\sB$, and where $\chi_a,\tilde\chi_a\in\Pi\Omega^{0,1}(\Sigma,\C^2)$ are fermionic gauge fields that impose the further constraints $\Omega^{\sA\sB}\rho_\sA\lambda_\sB^a=0$ and $\Omega^{\sA\sB}\tilde\rho_\sA\lambda_\sB^a=0$. These constraints couple $\rho$ and $\tilde\rho$ to the twistors and provide a supersymmetric extension of the bosonic worldsheet gauge algebra $\mathfrak{sl}_2\times \C$ for $S_{\scalebox{0.6}{$5d$}}^{\scalebox{0.6}{$\cN$}}$. Specifically, if we couple $S_{\scalebox{0.6}{$5d$}}^{\scalebox{0.6}{$\cN$}}$ to a total  of $p=1$ or $2$ independent copies of the $\rho\tilde\rho$ system, each with their own constraints,  the OPEs 
\begin{equation}
\begin{aligned}
&\left(\lambda^a_\sA\rho_r^\sA\right)\!(z)\, \left(\lambda^b_\sB \tilde \rho_s^\sB\right)\!(w) \sim \delta_{rs}\,\frac{\varepsilon^{ab}\,\la\lambda_\sA\lambda^\sA\ra}{z-w}\,,\\
&\left(\lambda^a_\sA\rho_r^\sA\right)\!(z)\, \langle\lambda^\sB\lambda_\sB\rangle(w)\sim 0\,,\qquad
\left(\lambda^b_\sA \tilde \rho_r^\sA\right)\!(z)\,\langle\lambda^\sB\lambda_\sB\rangle(w)\sim 0
\end{aligned}
\end{equation}
identify the worldsheet gauge algebra as $\mathfrak{sl}_2\ltimes H(0,2p)$, where $H$ denotes the Heisenberg Lie superalgebra\footnote{The Heisenberg superalgebra   $H(m_b,m_f)$ has a central element $z$, as well as $2m_b$ even and $m_f$ odd generators, $H=\la x_1,\dots, x_{2m_b},z \ra\oplus \la \psi_1,\dots, \psi_{m_f}\ra$. The generators satisfy the `usual' commutation relations
\[
 [x_i,x_{2i}]=z\,,\qquad \{\psi_r,\psi_s\}=2\delta_{rs}\,z\,.
\]}. 

Note that the form of these fermionic currents prevented us from writing an analogous system in 6d. While we could have chosen to define (say) $\tilde\rho^\sA$ as a section of dual spin bundle $\mathbb{S}^\sA$, allowing the kinetic term and one of the constraints in~\eqref{eqn:rho-action} to be written without of the help of $\Omega^{\sA\sB}$, the remaining term $\chi_a\, \lambda^{a}_{\sA}\rho^\sA$ either requires $\Omega^{\sA\sB}$ to raise the index on $\rho_\sA$, or else would have to couple to antichiral supertwistors in 6d, which are not present in the chiral model discussed above.

From the above components, we can construct three critical\footnote{Up to a central charge anomaly that agrees with the reduction from 10d, and can be cancelled by including a currents algebra encoding the 5 reduced dimensions.} models in 5d,
\begin{subequations}\label{eq:action_5d}
 \begin{align}
  &\text{Maximal Supergravity:} && S^{\scalebox{0.7}{sugra}}=S_{\scalebox{0.6}{$5d$}}^{\scalebox{0.6}{$\mathcal{N}=4$}}+S_{\rho_1}+S_{\rho_2}\\
  & \text{Maximal Super Yang-Mills:}  && S^{\scalebox{0.7}{{sYM}}}\,=S_{\scalebox{0.6}{$5d$}}^{\scalebox{0.6}{$\mathcal{N}=2$}}+\,S_{\rho}\;+S_C \\
  & \text{Bi-adjoint Scalar:} && S^{\scalebox{0.7}{BS}}\;\;\,=S_{\scalebox{0.6}{$5d$}}^{\scalebox{0.6}{bos}}\;\;\,+S_{C_1}+S_{C_2}\,,
 \end{align}
\end{subequations}
where $S_C$ denotes some internal current algebra. We briefly note that various other theories are possible using the worldsheet matter models of \cite{Casali:2015vta}, but they look less natural since they mix vector and spinor representations. For the remainder of this paper, we will focus on the models for maximal supergravity and maximal super Yang-Mills.

\subsection{BRST quantization and anomaly cancellation in 5d}  

Following the standard BRST procedure, we introduce  fermionic ghosts $\left\{(b,c), (m_{ab},n_{ab}), (m,n)\right\}$ associated to $\{e,A_{ab},a\}$ and bosonic ghosts 
 $\left\{(\beta _r^a,\gamma_r^a),(\tilde{\beta}_r^a,\tilde \gamma_r^a)\right\}$ 
 corresponding to $\{\chi_r^a,\tilde \chi_r^a\}$, respectively. Here, $r=1,2$ labels the two $\rho\tilde\rho$-systems in the gravity case, whereas for super Yang-Mills $r=1$. Each ghost system is chiral, e.g. $m_{ab},m\in\Pi\Omega^0(\Sigma,K_\Sigma)$ and  $\beta_r^a,\tilde\beta_r^a\in\Omega^0(\Sigma,K_\Sigma)$, whilst the familiar $bc$-ghost system has $c\in\Pi\Omega^0(\Sigma, T_\Sigma)$. As in string theory, it will be convenient to gauge fix $e=0$, and correlators include an integral over the moduli of the (marked) Riemann sphere. In the absence of vertex operators, we can similarly gauge $A_{ab}=a=\chi_a=\tilde\chi_a=0$, which gives a linear free gauge-fixed theory.  
 
The BRST operator takes the form 
\begin{equation}
Q=\oint c\left( T+\frac{1}{2}T^{\scalebox{0.7}{$bc$}}\right) + n_{ab}\left(J^{ab} +\frac12 m^a_{\ c}n^{bc}\right)+ n\la \lambda_{\sA}\lambda^{\sA} \ra  + \sum_r \gamma_{ra}\lambda^a_\sA\rho_r^\sA + \tilde{\gamma}_{ra} \lambda^a_\sA\tilde\rho_r^\sA+  m\gamma_{ra}\tilde\gamma^a_r\, ,
\end{equation}
where 
\begin{subequations}
\begin{equation}
\begin{aligned}
T &= \frac{1}{2}\mathcal{Z}_a\cdot \partial \mathcal{Z}^a  +\sum_r \left(\frac{1}{2}\tilde\rho_r^\sA \partial \rho_{\sA r} +\frac{1}{2}\rho_{\sA r}\partial\tilde\rho_r^{\sA}+ \beta_{ar}\partial\gamma_r^a  +\tilde\beta_{ar}\partial\tilde\gamma^a_r\right) \\
&\qquad + m^{ab}\partial n_{ab} + m\partial n
\end{aligned}
\end{equation}
is the matter holomorphic stress tensor and
\begin{equation}
J^{ab}=\frac{1}{2}\cZ^a\cdot \cZ^b +\sum_r \gamma_r^{(a}\beta_r^{b)}+\tilde\gamma_r^{(a}\tilde \beta_r^{b)}
\end{equation}
\end{subequations}
the SL$(2,\C)$ current, respectively.  

\medskip

As in 6d, anomalies in the worldsheet gauge and Virasoro algebras could obstruct $Q^2=0$.  Let us consider both cases in turn.  We first observe that the OPE between any current of the Heisenberg superalgebra and any other term in the BRST operator has at most a simple pole. Thus the Heisenberg superalgebra is always anomaly free.  The fields that transform in non-trivial representations of the SL$(2,\C)$ gauge algebra are the twistor superfields $\cZ^a=(\lambda_{\sA}^a,\mu^{\sA a},\eta_{\sI}^a)$ and the ghosts $(m_{ab},n_{ab})$ and  $\{(\beta_a, \gamma_a),(\tilde \beta_a,\tilde  \gamma_a)\}$. The ambitwistor string for {sYM} has one copy of the $\rho\tilde\rho$ system, so only one set of $\beta\gamma$ and $\tilde\beta\tilde\gamma$ ghosts. Its SL$(2,\C)$ anomaly coefficient is 
\begin{equation}
 \mathfrak{a}_{\scalebox{0.7}{$\mathrm{SL}(2)$}}^{\scalebox{0.6}{{sYM}}}
 =\sum_{i}(-1)^{F_i}\tr_{R_i}(t^kt^k)=\left(\frac{1}{2}\left(8-2\cN\right)
 +2
 \right)\tr_{\scalebox{0.7}{F}}(t^kt^k)-\tr_{\scalebox{0.7}{adj}}(t^kt^k)=\frac{3}{2}\left(2-\mathcal{N}\right)\,.
\end{equation}
Here, the factor of $\frac{1}{2}$ in front of the supertwistors is again due to their self-conjugacy, and the last equality follows from \eqref{eq:tr_sl2}. Thus the SL$(2,\C)$ anomaly vanishes in this model iff  $\mathcal{N}=2$, corresponding to maximally supersymmetric Yang-Mills theory in 5d. On the other hand, the ambitwistor string for supergravity has two sets of $\beta\gamma$ and $\tilde\beta\tilde\gamma$ ghosts, so its anomaly coefficient is
\begin{equation}
 \mathfrak{a}_{\scalebox{0.7}{$\mathrm{SL}(2)$}}^{\scalebox{0.6}{sugra}}
 =\left( 8-\mathcal{N}\right) \tr_{\scalebox{0.7}{F}}(t^kt^k)-\tr_{\scalebox{0.7}{adj}}(t^kt^k)=\frac{3}{2}\left(4-\mathcal{N}\right)\,.
\end{equation}
This vanishes iff $\cN=4$, corresponding to maximal supergravity in 5d.

Turning to the Virasoro anomaly, the central charge for the 5d {sYM} ambitwistor string is given by
\begin{equation}\label{eq:Virasoro5d_sYM}
 \mathfrak{c}^{\scalebox{0.6}{{sYM}}}=\left(-8+2\mathcal{N}\right)_{\scalebox{0.7}{$\mathcal{Z}$}} -26_{\scalebox{0.7}{$bc$}}+4\times(-2)_{\scalebox{0.7}{$mn$}}+4_{\scalebox{0.7}{$\rho\tilde\rho$}}+4\times 2_{\scalebox{0.7}{$\beta\gamma$}}+\mathfrak{c}_j=-30 +2\mathcal{N}+\mathfrak{c}_j\,.
\end{equation}
Since $\cN=2$ to cancel the gauge anomaly, this is critical for $\mathfrak{c}_j=10+16$, where the first 10 are obtained from compactifying 5 dimensions of an underlying 10d space-time, and an internal current algebra provides the remaining $\mathfrak{c}_j=16$.  For the supergravity model, we find similarly
\begin{equation}\label{eq:Virasoro5d_sugra}
 \mathfrak{c}^{\scalebox{0.6}{sugra}}=\left(-8+2\mathcal{N}\right)_{\scalebox{0.7}{$\mathcal{Z}$}} -26_{\scalebox{0.7}{$bc$}}+4\times(-2)_{\scalebox{0.7}{$mn$}}+2\left(4_{\scalebox{0.7}{$\rho\tilde\rho$}}+4\times 2_{\scalebox{0.7}{$\beta\gamma$}}\right)=-18 +2\mathcal{N}\,.
\end{equation}
Since $\mathcal{N}=4$, we obtain a critical model if we again include a CFT of central charge $\mathfrak{c}_{\scalebox{0.7}{5d}}=10$ arising from five compactified dimensions.

\subsection{Vertex operators for supermultiplets}
\label{sec:5dvertex}

Unlike the RNS ambitwistor string,  the vertex operators for {sYM} and sugra states in our twistor representation of the ambitwistor string will have manifest target space supersymmetry. Let us briefly recall the structure of on-shell, linearised supermultiplets in these theories, following~\cite{Cheung:2009dc,Dennen:2009vk,Huang:2010rn,Cachazo:2018hqa,Heydeman:2017yww}.  

The supersymmetry algebra $\{Q_{\sA \sI},Q_{\sB \sJ}\}=2\,\omega_{\sI\sJ}P_{\sA\sB} $ means the supercharges may be represented as $Q_{\sA\sI}=\kappa_\sA^a Q_{a\sI}$ when acting an on-shell massless supermultiplet with momentum $k_{\sA\sB}=\la\kappa_\sA\kappa_\sB\ra/2$. The on-shell supercharges $Q_{a\sI}$ transform in the bi-fundamental of ${\rm SL}(2)\times{\rm Sp}(2\cN)$ and obey
\begin{equation}
\label{on-shell-Qs}
	\left\{Q_{a\sI},Q_{b\sJ}\right\} = \varepsilon_{ab}\,\omega_{\sI\sJ}\,.
\end{equation}
Representations of this on-shell algebra can keep either the little group or the R-symmetry group manifest, but not both. In this paper, we will use the R-symmetry preserving representation\footnote{This choice is motivated in part by the fact that the $\epsilon_{a}$ needed to write the bosonic plane wave already spoils manifest little-group invariance. In fact,~\eqref{mom-estate}  already provides a basis $\{\epsilon_a,v_a\}$ of this space. However, for each insertion point, the value of $v_{a}$ depends on the solution of the polarized scattering equations and it is convenient to avoid this by introducing $\xi_a$. See \cite{Albonico:2020mge} for more details.}. We thus choose a basis $\{\epsilon_a,\xi_a\}$ (with $\la\epsilon\xi\ra=1$) of the 2-dimensional representation of SL(2) and represent
\begin{equation}
\label{eq:def_susy-gen}
Q_{a\sI}=\left(\xi_a q_\sI +\epsilon_a\,\omega_{\sI\sJ}\frac{\p}{\p q_\sJ}\right) 
\end{equation}
when acting on functions $\tilde\F(\kappa_{a\sA},q_\sI)$ on our on-shell momentum superspace. In particular, in this representation the sYM multiplet is described by an on-shell superfield
\[
\begin{aligned}
\tilde\F(\kappa,q) &= (\xi_a\xi_b + q^2\epsilon_{(a}\xi_{b)} + q^4\epsilon_a\epsilon_b)F^{ab}(\kappa) + (\xi_a+\epsilon_a q^2)q_\sI\Psi^{a\sI}(\kappa) + (q_\sI q_\sJ - \frac{1}{4}\omega_{\sI\sJ}q^2)\phi^{\sI\sJ}(\kappa)\\
	&= F^{\epsilon\epsilon}(\kappa) + q_\sI\Psi^{\epsilon\sI}(\kappa) + q^2 F^{\epsilon\xi}(\kappa) + \frac{1}{2}q_\sI q_\sJ\phi^{\sI\sJ}(\kappa) +  q^2 q_{\sI} \Psi^{\xi\sI}(\kappa)+ q^4 F^{\xi\xi}(\kappa)
\end{aligned}
\]
in momentum superspace, where $F^{ab}(\kappa)=F^{(ab)}(\kappa)$ are the three on-shell polarizations of the gluon, $\Psi^{a\sI}(\kappa)$ are gluinos, and $\phi^{\sI\sJ}(\kappa)=\phi^{[\sI\sJ]}(\kappa)$ are five scalars ($\omega_{\sI\sJ}\phi^{\sI\sJ}(\kappa)=0$).
 
We note that
\begin{equation}
\F_{(\kappa,q)}(x,\theta)=\exp\left(\frac{i}{2} x^{\sA\sB}\la\kappa_\sA\kappa_\sB\ra  +\frac{i}{2} \la \xi \kappa_\sA\ra\, \left( \omega_{\sI\sJ}\theta^{\sI\sA}\theta^{\sJ\sB} \la \kappa_\sB\epsilon\ra+q_\sI\theta^{\sI\sA}\right)\right)
\label{susy-plane-wave}
\end{equation}
obeys the interwining relation
\begin{equation}
\label{eq:intertwining}
	\left(\frac{\p}{\p\theta^{\sI\sA}} + \omega_{\sI\sJ}\theta^{\sJ\sB}\frac{\p}{\p x^{\sA\sB}}\right)\F_{(\kappa,q)}(x,\theta) =\frac{i}{2} \left(q_\sI\la\xi\kappa_\sA\ra +\epsilon_\sA \omega_{\sI\sJ}\frac{\p}{\p q_\sJ}\right)\F_{(\kappa,q)}(x,\theta)\,,
\end{equation}
where 
\[
Q_{\sA\sI}=\frac{\p}{\p \theta^{\sA\sI}}+\omega_{\sI\sJ}\theta^{\sJ\sB}\frac{\p}{\p  x^{\sA\sB}}
\] 
is the supercharge acting on Minkowski superspace. Thus~\eqref{eq:intertwining} is the correct supersymmetric generalization of a plane wave\footnote{Note that $\omega_{\sI\sJ}\theta^{\sI\sA}\theta^{\sJ\sB}$ is symmetric in $AB$, so the quadratic term in $\theta$s in the exponential of~\eqref{susy-plane-wave} cannot be treated as part of $x^{\sA\sB}$.}. See \cite{Mason:2012va,Geyer:2018xgb, Albonico:2020mge} for more details on this supersymmetry representation.

\medskip

We now construct vertex operators based on this supersymmetric plane wave. As in 6d, these can be understood as the supersymmetric ambitwistor Penrose transform, relating  linear fields of spin $s$ on space-time to Dolbeault cohomology classes  $ H^1(\mathbb{PA}_{\scalebox{0.7}{$5|2\mathcal{N}$}},\mathcal{O}(2s-2))$ on ambitwistor space~\cite{Mason:2012va}.  On ambitwistor space, the supercharge is represented as
\[
Q_{\sA\sI} = \lambda^a_{\sA}\frac{\p}{\p \eta^{a\sI}}+\omega_{\sI\sJ}\eta^{a\sJ}\frac{\p}{\p\mu^{a\sA}}
\]
Following \cite{Geyer:2018xgb, Albonico:2020mge}, the supersymmetric extension of the basic vertex operator \eqref{mom-estate} is given by 
\begin{equation}
V =\int d^2u \, d^2v  \;w(u)\;\bar\delta\big(\langle v  \epsilon  \rangle -1 \big)\,\bar \delta^4\Big(\langle u \lambda_\sA\rangle -\langle v \kappa_{\sA}\rangle\Big)\,\e^{ i\la u  \mu^{\sA}\ra \epsilon_{\sA}+\la u \eta^\sI\ra q_{\sI} -\frac{1}{2}\la \xi v \ra q^2}\, ,\label{super-mom-estate}
\end{equation}
where the exponential factor obeys an intertwining relation between $\P\A{\scalebox{0.7}{$5|2\mathcal{N}$}}$ and on-shell momentum superspace, as in~\eqref{eq:intertwining}.
Again, the choice of quadratic differential $w$ in~\eqref{super-mom-estate} determines the state (or supermultiplet) in the target space, and is theory dependent. Here, we allow $w$ to depend on the parameter $u$ as well as the worldsheet matter systems.

A crucial feature of the vertex operator~\eqref{super-mom-estate} is that it is already invariant under half of the Heisenberg superalgebra.  Specifically, the integrand of $V$ is invariant under the currents $\la u\lambda_\sA\ra \rho^\sA$ and $\la u\lambda_{\sA}\ra\tilde\rho^\sA$, provided these currents have trivial OPEs with the quadratic differential $w$. Therefore, to obtain a BRST-closed vertex operator it is sufficient to include $\delta$-functions that force only the ghost components orthogonal  to $u$ to vanish. Consequently, vertex operators for supermultiplets (with `picture number' $(-1,-1)$) in {sYM} and supergravity are given by~\eqref{super-mom-estate} with the choices
\begin{equation}\label{eqn:w-Sugra}
\begin{aligned}
 w_{\scalebox{0.6}{{sYM}}}^{\scalebox{0.6}{$(-1,-1)$}} &= n\,\delta(\left\la u\gamma\ra\right)\delta(\left\la u\tilde \gamma\ra\right)\;t_{\mathfrak{a}}j^{\mathfrak{a}}\\
 w_{\scalebox{0.6}{sugra}}^{\scalebox{0.6}{$(-1,-1)$}} &= n\,\prod_{r=1,2}\delta(\left\la u\gamma_r\ra\right)\delta(\left\la u\tilde \gamma_r\ra\right)
\end{aligned}
\end{equation}
in the heterotic  and Type II ambitwistor strings, respectively. Here, $n$ denotes the ghost associated to the current $\la \lambda_\sA\lambda^\sA\ra$ that reduces the theory to 5d. Note that, since $u$ transforms as a section of $T^{\scalebox{0.6}{$1/2$}}_{\scalebox{0.6}{$\Sigma$}}$ (as dictated by the exponential in~\eqref{super-mom-estate}) whilst all the $\gamma$ ghosts are worldsheet scalars,  $ w_{\scalebox{0.6}{{sYM}}}$ and $ w_{\scalebox{0.6}{sugra}}$ are indeed quadratic differentials. Including the standard $bc$ ghost system, $cV$ and $\int_\Sigma V$ are well-defined fixed and integrated vertex operators, respectively.

We remark that, as in \cref{sec:6d_VO}, the heterotic  theory contains further states corresponding to a  higher-order 5d (super)conformal gravity. Vertex operators for these states follow from choosing
\begin{equation}\label{eqn:w-{sYM}}
 w_{\scalebox{0.6}{cgrav}}^{\scalebox{0.6}{$(-1,-1)$}} = n\,\delta(\left\la u\gamma\ra\right)\delta(\left\la u\tilde \gamma\ra\right)\;E_{\scalebox{0.6}{$\mathcal{A}\mathcal{B}$}}\cZ^{\scalebox{0.6}{$\mathcal{A}$}}_a\cZ^{\scalebox{0.6}{$\mathcal{B}$}a}\,,
\end{equation}
where the polarization tensors $E_{\scalebox{0.6}{$\mathcal{A}\mathcal{B}$}}$ are in the Lie algebra of the superconformal group and have to satisfy a supersymmetric extension of the relations \eqref{eq:conf_cons_rels} for the vertex operator to be well-defined. These states are absent in the Type II model, as they are not invariant under both copies ($r=1,2$) of the worldsheet Heisenberg algebra.

\subsection{Moduli and picture changing operators}
\label{sec:PCOs}

The  factors of $n$ and $\delta(\la u\gamma\ra) \,\delta(\la u\tilde\gamma\ra)$ in the vertex operators constrain these fermionic and bosonic ghosts to vanish at these insertion points. Consequently, the corresponding gauge fields will have moduli which cannot be set to zero by a BRST transformation.

In the bosonic case, the moduli of the gauge field $a$  associated to $\la\lambda_\sA\lambda^\sA\ra$ live in the cohomology group $H^{0,1}(\Sigma,\cO(-\sigma_1-...-\sigma_n))$ of worldsheet $(0,1)$-forms that cannot be written as $\bar\p g$ for any function $g$ that vanishes at the insertion points.  Picking a basis $\{h_j\}$ of representatives of this space, at genus zero we always impose a gauge in which $a$ takes the form
\begin{equation}
 a^{\scalebox{0.6}{GF}}=\sum_{j=1}^{n-1}a_j h_j
\end{equation}
for some choice of bosonic parameters $a_j$. We let $\alpha_j = \{Q,a_j\}$ be the BRST transformation of these parameters and let $P_5 = \{Q,m\}$ be the Nakanishi-Lautrup field associated to the antighost $m$. Then the ghost and gauge-fixing terms for $a$ are
\begin{equation}
\begin{aligned}
 S_{\scalebox{0.7}{GF}}&= \int_\Sigma\left\{Q,\,m(a-a^{\scalebox{0.6}{GF}})\right\}\\
 &= \int_\Sigma m\,\bar\p n + P_5(a -a^{\scalebox{0.6}{GF}}) - \sum_{j=1}^{n-1}\alpha_j\int_\Sigma m\, h_j
\end{aligned}
\end{equation}
Integrating out $P_5$ enforces the gauge-fixing condition $a= a^{\scalebox{0.6}{GF}}$ so that the coupling of $a$ to its current becomes
\[
	\int_\Sigma a\la\lambda_\sA\lambda^\sA\ra = \sum_j a_j \int_\Sigma h_j \la\lambda_\sA\lambda^\sA\ra \,.
\]
The further integrals over the moduli parameters $a_j$ and $\alpha_j$ lead to $n-1$ insertions of the form\footnote{See also the discussion in e.g. \cite{Witten:2012bh,Adamo:2013tsa}.}
\begin{equation}\label{eq:delta_fns_a}
 \Xi_j=\left(\int_\Sigma h_j\, m \right)\;\delta\!\left(\int_\Sigma h_j\, \la \lambda_{\sA}\lambda^{\sA} \ra \right)\,,
\end{equation}
in the correlator. It will be convenient to choose the basis $h_j$ to extract the residue at the marked point $\sigma_j$. The $n-1$ insertions of the antighost $m$ then absorb $n-1$ factors of the insertion of the $n$-ghost, with the one remaining $n$-ghost insertion saturating its zero mode integral. The bosonic $\delta$-functions in $\Xi_j$ help implement the reduction from six to fix dimensions in the presence of vertex operators, as we see in more detail below.

\medskip

We now consider the moduli of the fermionic gauge fields $\chi^a$ and $\tilde\chi^a$. As discussed above, the vertex operators built from~\eqref{eqn:w-Sugra} only constrain one component of each ghost to vanish at each insertion point. More specifically, for a (local) basis $u$ and $\hat u$ with $\la u \hat u\ra\neq 0$, the vertex operators force the ghost components  $\gamma_{\hat u}=\la u\gamma\ra$ and $\tilde\gamma_{\hat u} = \la u\tilde\gamma\ra$ to vanish, but leave $\gamma_u=\la \hat u \gamma\ra$ and $\tilde\gamma_u = \la\hat u \tilde \gamma\ra$ unconstrained. Correspondingly, even in the presence of vertex operators, we are free to use BRST transformations to set $\chi_u=\la \hat u\chi\ra=0$, whilst $\chi_{\hat u}=\la u\chi\ra$ develops moduli (and similarly for $\tilde\chi$). This argument applies at each marked point $\sigma_i$. Globally, it will be convenient to pick a $\hat u$, transforming as a section of $K^{\scalebox{0.6}{$1/2$}}_{\scalebox{0.6}{$\Sigma$}}$ and satisfying $\la u_i\hat u\ra\neq 0$ for all $u_i$, such that $\{u_i,\hat u\}$ provides a (local) basis of $\C^2$ at each marked point. The gauge-fixing term then becomes
\begin{equation}\label{eqn:chi-gauge-fix}
 S_{\scalebox{0.7}{GF}}=\int\left\{Q,\,\beta^a(\chi_a-\chi_a^{\scalebox{0.6}{GF}})+\tilde \beta^a(\tilde \chi_a-\tilde \chi_a^{\scalebox{0.6}{GF}})\right\}\,,
\end{equation}
where the gauge-fixed $\chi_a^{\scalebox{0.6}{GF}}$ and $\tilde \chi_a^{\scalebox{0.6}{GF}}$ are given by
\begin{equation}
 \chi^{\scalebox{0.6}{GF}}_a =\hat u_{ a} \sum_{l=1}^{n-2}\chi_{l} \,h^{\scalebox{0.6}{$\chi$}}_l\,,
 \qquad \tilde \chi^{\scalebox{0.6}{GF}}_a = \hat u_{ a}\sum_{l=1}^{n-2}\tilde \chi_{l}\, h^{\scalebox{0.6}{$\chi$}}_l\,.
\end{equation}
This choice means $\la\hat{u}\chi^{\scalebox{0.6}{GF}}\ra =0$  whilst $\chi^{\scalebox{0.6}{GF}}_{\hat u}=\la u\chi^{\scalebox{0.6}{GF}}\ra$ is expanded in terms of a basis $\{h_l^\chi\}$ of\footnote{Since the field $\chi\in\Omega^{0,1}(\Sigma,K_\Sigma)$ whilst $u\in T^{\scalebox{0.6}{$1/2$}}_{\scalebox{0.6}{$\Sigma$}}$, the component $\chi_{\hat u} = \la u \chi\ra \in\Omega^{0,1}(\Sigma,K^{\scalebox{0.6}{$1/2$}}_{\scalebox{0.6}{$\Sigma$}})$. The cohomology group $H^{0,1}(\Sigma,K^{\scalebox{0.6}{$1/2$}}_{\scalebox{0.6}{$\Sigma$}}(-\sigma_1-...-\sigma_n))$ describing the moduli of $\chi_{\hat u}$ has dimension $n-2$.}  $H^{0,1}(\Sigma,K^{\scalebox{0.6}{$1/2$}}_{\scalebox{0.6}{$\Sigma$}}(-\sigma_1-...-\sigma_n))$ with fermionic coefficients $\chi_l$ (and similarly for $\tilde\chi$). Following an argument equivalent to that used for the moduli of $a$,  integration over the expansion parameters $\chi_l$ and $\tilde\chi_l$ leads to  $2\times (n-2)$ `picture changing operator' (PCO) insertions in the correlator,
\begin{equation}\label{eq:PCO}
 \Upsilon(z_l)=\delta(\la \hat u \beta\ra )\, \la \hat u\lambda_\sA\ra \rho^\sA\,,\qquad \tilde{\Upsilon}(z_l) =\delta(\la \hat u \tilde\beta\ra )\, \la \hat u\lambda_\sB\ra\tilde \rho^\sB\,.
\end{equation}
Again, these picture changing operators absorb $(n-2)$ of the ghosts $\gamma$ in the vertex operators, leaving two factors to saturate the remaining zero-modes (and similarly for $\tilde\gamma$).

Above, we have chosen the basis $h^{\scalebox{0.6}{$\chi$}}_l$ such that it evaluates the respective fields at $n-2$ reference points $z_l$. Since the $h^{\scalebox{0.6}{$\chi$}}_l$ originate in the BRST-exact term~\eqref{eqn:chi-gauge-fix}, the final correlator will be independent of the specific choice of $z_l$. As usual, we can use this freedom to define vertex operators with different `picture' numbers, either by choosing the basis $h^{\scalebox{0.6}{$\chi$}}_l$ to extract the residue at marked points $\sigma_l$, or via the limit
\begin{align}
 w^{\scalebox{0.6}{$(-1,0)$}} (\sigma_i) &= \mathrm{lim}_{z_l\rightarrow \sigma_i}\tilde \Upsilon (z_l)\, w^{\scalebox{0.6}{$(-1,-1)$}}(\sigma_i) \,.
\end{align}
For example, the {sYM} multiplet  has picture $(-1,0)$ and $(0,0)$ vertex operators built from 
\begin{equation}
 w_{\scalebox{0.6}{{sYM}}}^{\scalebox{0.6}{$(-1,0)$}} = n\,\delta(\left\la u\gamma\ra\right)\,\epsilon_\sA \tilde\rho^\sA \;t_{\mathfrak{a}}j^{\mathfrak{a}}\qquad\text{or}\qquad
 w_{\scalebox{0.6}{{sYM}}}^{\scalebox{0.6}{$(0,0)$}} = n\,\left( \frac{\la \hat u\lambda_\sA\ra\epsilon^\sA}{\la u\hat u\ra} + \epsilon^\sA\epsilon_\sB\, \rho_\sA \tilde\rho^\sB\right) \;t_{\mathfrak{a}}j^{\mathfrak{a}}
\end{equation}
together with the plane wave  in~\eqref{super-mom-estate}. We will see below that, after performing the path integral over the twistor variables,  $w_{\scalebox{0.6}{{sYM}}}^{\scalebox{0.6}{$(0,0)$}}$ is independent of $\hat u$. The first term simplifies then to $-e\cdot \la\lambda\lambda\ra$, where $e$ is the polarization vector of the gluon. Analogous descent formul{\ae} give the vertex operators in various pictures for the $\cN=4$ supergravity multiplet in the Type II model (and the $\cN=2$ conformal gravity supermultiplet in the heterotic model).


\subsection{Correlators and amplitudes}\label{sec:amplitudes}
After integrating over the moduli of the worldsheet gauge fields, $n$-point amplitudes arise from the   correlator
\begin{equation}\label{eq:corr_gen}
 \mathcal{M}_n=\left\langle cV_1^{\scalebox{0.6}{$(-1,-1)$}}\,cV_2^{\scalebox{0.6}{$(-1,-1)$}}\,cV_3^{\scalebox{0.6}{$(-1,-1)$}} \prod_{i=4}^n \int_\Sigma V_{i}^{\scalebox{0.6}{$(-1,-1)$}}\; \prod_{j=1}^{n-1}\Xi_j\;\prod_{l=1}^{n-2}\Upsilon_l\tilde{\Upsilon}_l \right\rangle\,.
\end{equation}
Inspecting the explicit form of the operators, we note that all dependence on the  $\mu^\sA_a$ components of twistors and the fermions $\eta$  comes from the exponentials $\e^{ i\la u_i \mu^{\sA}\ra \epsilon_{i\sA}+\la u_i\eta^\sI\ra q_{i\sI}}$ in the vertex operators. We can use this simple dependence to perform the path integral over the $\mathcal{Z}$-system by including these terms into the action as sources. The path integral then localizes onto the classical equations of motion,
\begin{equation}
 \dbar \lambda_\sA^a = \sum_{i=1}^n u_i^a\epsilon_{i\sA}\,\bar\delta(\sigma-\sigma_i)\,, \qquad
  \dbar\eta_\sI^a =\frac{1}{2} \sum_{i=1}^n u_i^aq_{i\sI}\,\bar\delta(\sigma-\sigma_i)\,,
\end{equation}
which are solved uniquely by
\begin{subequations}
\begin{equation}\label{eq:sol_lambda}
 \lambda_\sA^a(\sigma) = \sum_{i=1}^n \frac{u_i^a\epsilon_{i\sA}}{\sigma-\sigma_i}\,,
 \qquad
 \eta_\sA^a(\sigma) = \frac{1}{2}\sum_{i=1}^n \frac{u_i^aq_{i\sI}}{\sigma-\sigma_i}
\end{equation}
at genus zero. Together with the normalization conditions $\la v_i\epsilon_i\ra=1$ and constraints
\begin{equation}\label{eqn:lambda-constraints}
	\la u_i\lambda_\sA(\sigma_i)\ra = \la v_i \kappa_{i\sA}\ra
\end{equation}
\end{subequations}
coming from the $\bar\delta$-functions in each vertex operator, equations~\eqref{eq:sol_lambda}-\eqref{eqn:lambda-constraints} will constitute the \emph{polarized scattering equations}. 

Localization to solutions of \eqref{eq:sol_lambda} has several important consequences for the amplitude formul{\ae}. First, it ensures that $\la u_i\lambda_\sA(\sigma)\ra$ is regular at $\sigma=\sigma_i$, so the remaining polarized scattering equations~\eqref{eqn:lambda-constraints} are well-defined. The same applies to the fermionic terms in the exponent. Substituting~\eqref{eq:sol_lambda} back into the vertex operators, the amplitudes now only depend on the supermomenta $q_i$ via the exponential $e^{F_n}$, where
\begin{equation}
 F_{\scalebox{0.7}{$n$}}=\sum_{i<j}   \frac{ \langle u_i u_j\rangle q_{iI}q_j^I}{\sigma_{ij}} - 
 \frac{1}{2}\sum_{i=1}^n \la \xi_iv_i\ra q_i^2\,.
\end{equation}
This guarantees that the amplitudes are supersymmetrically invariant \cite{Albonico:2020mge}.

Localization onto \eqref{eq:sol_lambda} also allows us to make contact with the RNS ambitwistor string and the spinorial resolution  $P_{\sA\sB}={\frac{1}{2}}\la \lambda_\sA\lambda_\sB\ra$ mentioned in \eqref{P-lambda}. In the RNS ambitwistor string, a very similar argument leads to \cite{Mason:2013sva}
\begin{equation}
 P_\mu^{\scalebox{0.6}{RNS}}=\sum_{i=1}^n\frac{k_{i\mu}}{\sigma-\sigma_i}d\sigma\,,
\end{equation}
and the scattering equations impose the conditions that $P^2=0$ on the sphere. In contrast, the ambitwistor string discussed here always resolves the null condition due to $P_{\sA\sB}={\frac{1}{2}}\la \lambda_\sA\lambda_\sB\ra$. Instead, the polarized scattering equations ensure that the the residues of $P$  are given by the external momenta,
\begin{equation}
\frac{1}{2} \mathrm{Res}_{\sigma_i}\la \lambda_\sA(\sigma)\lambda_\sB(\sigma)\ra =\frac{1}{2}\epsilon_{[\sA} \la u_i \lambda_{\sB]}(\sigma_i)\ra=k_{i[\sA\sB]}\,,
\end{equation}
and thus indeed $P= P^{\scalebox{0.6}{RNS}}$.

At this point, let us revisit the integrated vertex operators. Using the explicit form of $\lambda_\sA^a(\sigma)$ in~\eqref{eq:sol_lambda}, it is easily checked \cite{Geyer:2018xgb} that
\begin{equation}
 \lambda_\sA^a(\sigma_i)\,\epsilon^\sA=u_i^a\,\big(e_i\cdot P(\sigma_i)\big)\,,
\end{equation}
where $e_i$ is the polarization vector. The `picture-zero' vertex operator thus simplifies to
\begin{equation}
 w_{\scalebox{0.6}{{sYM}}}^{\scalebox{0.6}{$(0,0)$}} = n\,\left(- e\cdot P+ \epsilon^\sA\epsilon_\sB\, \rho_\sA \tilde\rho^\sB\right) \;t_{\mathfrak{a}}j^{\mathfrak{a}}
\end{equation}
for sYM. The localization also suggests a convenient route for performing the path integral over the  $mn$ system. If we choose $\Xi_j$ with a basis $h_j$ that extracts the residue at the  marked point $\sigma_{j}$, the $mn$ path integral becomes trivial and accounting for the moduli $a_j$ of $a$ reduces to an insertion of 
\begin{equation}\label{eq:delta_5d}
 \prod_{j=1}^{n-1}\delta\left(\mathrm{Res}_{\sigma_i}\la\lambda_\sA\lambda^\sA\ra\right)=\prod_{j=1}^{n-1}\delta(\Omega\cdot k_{j})\,.
\end{equation}
These have a very intuitive physical interpretation in restricting the momenta of $n-1$ particles to five dimensions --- the expected implication of quantizing the constraint $\la\lambda_\sA\lambda^\sA\ra$ enforcing the reduction in the action.  The remaining constraint for the $n^{\mathrm{th}}$ particle follows from the  momentum conservation  encoded in the  polarized scattering equations.

Finally, accounting for the quotient by SL$(2,\mathbb{C})_\sigma\times\mathrm{SL}(2,\mathbb{C})_u$ arising from the path integrals over the  $bc$ and $m_{ab}n^{ab}$-systems, the correlator becomes
\begin{equation}\label{eq:ampl_susy}
 \mathcal{M}_n=\int \rd\mu_n^{\mathrm{pol}}\ \mathcal{I}_n\;e^{F_N}\,.
\end{equation}
Here, $\rd\mu_n^{\mathrm{pol}}$ is the 5d measure composed of the polarized scattering equations of the vertex operators,  as well as the $\delta$-functions \eqref{eq:delta_5d} responsible for the reduction to 5d:
\begin{equation*}
 \rd\mu_n^{\mathrm{pol}}= \frac{\prod_l \rd \sigma_l\,\rd^2 u_l\,\rd^2 v_l}{\mathrm{vol}\; \mathrm{SL}(2,\mathbb{C})_\sigma \times \mathrm{SL}(2,\mathbb{C})_u}\, \prod_{j=1}^{n-1}\delta(\Omega \cdot k_{j})\,\prod_{i=1}^n  \bar\delta\big(\langle v_i \varepsilon_i \rangle -1 \big)\;\bar\delta^4\Big(\langle u_{i}\lambda_\sA(\sigma_i)\rangle -\langle v_{i}\kappa_{i\sA}\rangle\Big)\,,
\end{equation*}
see also \cref{polscatt} for details. The form  \eqref{eq:ampl_susy} is universal, following only from the bare structure of the vertex operators and the ambitwistor string action.

All theory-dependence resides in the integrand $\mathcal{I}_n$, calculated from the path integral over the worldsheet matter systems,
\begin{equation}
  \mathcal{I}_n=\left\langle  w_{1}^{\scalebox{0.6}{$(-1,-1)$}} w_{2}^{\scalebox{0.6}{$(-1,-1)$}} \prod_{i=3}^n w_{i}^{\scalebox{0.6}{$(0,0)$}}\right\rangle\,.
\end{equation}
on support of the localization for  $\lambda_\sA^a$.

Let us  evaluate the $\mathcal{I}_n$ for either $n$ external super Yang-Mills multiplets in the heterotic model, or $n$ external $\cN=4$ supergravity multiplets in the Type II model. The internal current algebras $j^{\mathfrak{a}}$ in sYM or for (each copy of) the Heisenberg superalgebra do not interact, so we can split the correlators further into `half-integrands', each arising from one of these matter systems. The simplest case is the current algebra correlator in the Yang-Mills model, giving the familiar Parke-Taylor factor
\begin{equation}
 \mathrm{PT}(\alpha)=\frac{\tr(t_{\alpha(1)}\dots t_{\alpha(n)})}{\sigma_{\alpha(1)\alpha(2)}\dots \sigma_{\alpha(n)\alpha(1)}}
\end{equation}
at leading trace\footnote{As usual, sub-leading trace contributions will arise from tree-amplitudes with external sYM states connected by conformal graviton propagators. We neglect these.}, summed over dihedrally inequivalent orderings $\alpha\in S_n/D_n$. The other half-integrand contains the correlators over the $\rho\tilde\rho$-system, as well as the associated  $\beta\gamma$ and $\tilde \beta\tilde\gamma$ ghost systems. The ghost correlators simply contribute 
\begin{equation}
 \Big\langle \delta(\langle u_1 \gamma(\sigma_1)\rangle)\, \delta(\langle u_2 \gamma(\sigma_2)\rangle) \Big \rangle_{\beta\gamma}=\ \Big\langle \delta(\langle u_1 \tilde \gamma(\sigma_1)\rangle)\, \delta(\langle u_2 \tilde \gamma(\sigma_2)\rangle) \Big \rangle_{\tilde \beta\tilde \gamma}=\ \frac{1}{\langle u_1 u_2\rangle}\,.
\end{equation}
 The correlators over the $\rho\tilde\rho$ system gives the determinant of an $(n-2)\times(n-2)$  matrix $H^{[12]}_{[12]}$, with entries indexed by $i,j\in\{3,4,\ldots,n\}$ given by
\begin{equation}
H_{ij} =\frac{\epsilon_{i\sA}\epsilon^\sA_j}{\sigma_{ij}}\,,\qquad H_{ii}=
 -e_i\cdot P(\sigma_i)\,.
\end{equation}
The full correlator thus combines to a \emph{reduced} determinant, 
\begin{equation}
 \left\langle  w_{1}^{\scalebox{0.6}{$(-1,-1)$}} w_{2}^{\scalebox{0.6}{$(-1,-1)$}} \prod_{i=3}^n w_{i}^{\scalebox{0.6}{$(0,0)$}}\right\rangle_{\substack{\rho\tilde\rho\;\;\;\;\;\\\beta\gamma,\tilde\beta\tilde\gamma}}=\frac{\det H^{[12]}_{[12]}}{\la u_1 u_2\ra^2}={\det}'H\,,
\end{equation}
that is  invariant under permutations of all particle labels. This permutation invariance can be established directly \cite{Geyer:2018xgb}, but is also manifest from the correlator \eqref{eq:corr_gen} using general picture changing operators; different removed rows and columns in the reduced determinant correspond to different basis choices in the picture changing operators, or different picture numbers of the vertex operators --- none of which affect the final amplitude.

Putting everything together, the integrands for $n$-particle states in super Yang-Mills and supergravity are
\begin{equation}
 \mathcal{I}_n^{\scalebox{0.7}{sugra}} = {\det}'H\ {\det}'\tilde H\,,\qquad \mathcal{I}_n^{\scalebox{0.7}{{sYM}}} = {\det}'H\sum_{\alpha\in S_n/D_n}\mathrm{PT}(\alpha)\,.
\end{equation}
The genus zero worldsheet correlator of the Type II ambitwistor string in the twistor representation thus reproduces the manifestly supersymmetric $n$-particle tree amplitude 
\begin{equation}
	\mathcal{M}_n^{\scalebox{0.6}{grav}} =\int \rd\mu_n^{\mathrm{pol}}\ {\det}'H\ {\det}'\tilde H\;e^{F_N}
\end{equation}
of maximal ($\cN=4$) supergravity in 5d, while the leading-trace  part of the correlator of $n$ SYM states in the heterotic ambitwistor string reproduces the $n$-particle tree amplitude
\begin{equation}
	\mathcal{M}_n^{\scalebox{0.6}{sYM}} =\sum_{\alpha\in S_n/D_n}\int \rd\mu_n^{\mathrm{pol}}\ {\det}'H\ \mathrm{PT}(\alpha) \;e^{F_N}
\end{equation}
of 5d maximal ($\cN=2$) sYM. These representations of the superamplitudes were first obtained in \cite{Geyer:2018xgb} and were improved to be presented in this form  in~\cite{Albonico:2020mge} where they were proved via BCFW recursion.

\section{Discussion}

In this paper we gave a presentation of ambitwistor strings in six and five dimensions based on twistors, i.e., the spinor representation of the conformal group.   In this description, the $P^2=0$ constraint of the RNS ambitwistor string is solved instead of gauged.  The string couples directly to vertex operators built out of the Penrose transforms of space-time fields, represented as cohomology classes on ambitwistor space (and indirectly from twistor space). By contrast, in the RNS model of the ambitwistor string, the Penrose transform from space-time to ambitwistor space is implemented by the descent arising from the BRST treatment of the gauging of the $P^2=0$  constraint \cite{Mason:2013sva,Adamo:2013tsa}.

In six dimensions we were only able to obtain an analogue of the bosonic ambitwistor string of \cite{Mason:2013sva}, with the additional worldsheet fields needed to describe ordinary (2$^{\rm nd}$ order) Yang-Mills and gravity remaining elusive. It was already well known from the work of \cite{Johansson:2017srf,Azevedo:2017lkz,Berkovits:2018jvm, Azevedo:2019zbn} that the bosonic ambitwistor string contains a gauge sector with a $(DF)^2$ Lagrangian and the corresponding double copy gravitational theory.  The novelty here is that the twistorial representation of the ambitwistor string makes the 6d conformal invariance of these theories manifest.  The difficulty with the analogue of the heterotic and type II models is that they would correspond to the incorporation of one or two copies of \eqref{eqn:rho-action} which requires ingredients of both helicities, $\lambda_\sA^a$ and $\tilde{\lambda}^\sA_{\dot a}$.  To get around this, it seems that one would have to consider a model incorporating both $Z^a$ and $\tilde Z_{\dot a}$.  would then need to have cumbersome constraints, essentially that $Z^1\wedge Z^2=\tilde Z_{\dot 1}\wedge \tilde Z_{\dot 2}$ in the sense that $Z^1\wedge Z^2$ and $\tilde{Z}_{\dot 1}\wedge\tilde{Z}_{\dot 2}$ should define the same element of $\mathfrak{so}(8)$ via triality.

In five dimensions,  we are able to construct fully fledged twistor representations of the heterotic and type II ambitwistor strings by incorporating one or two copies of the $\rho\tilde\rho$-system \eqref{eqn:rho-action}. It is interesting to compare this to the situation in 4d.  In 4d, there is no additional current algebra for YM amplitudes, but there is for gravity.  In fact, it is easy to see the close resemblance of the $\rho\tilde\rho$-system to that in  \cite{Skinner:2013xp} for the 4d $\mathcal{N}=8$ twistor string. However the contribution of the second $\rho\tilde\rho$-system  is absorbed essentially in the change of variables from the CHY measure, which is the same as the polarized scattering equations measure, and that of the twistor string or the 4d ambitwistor-string. Given its prevalence in spinorial worldsheet models  for Yang-Mills theory and gravity, it would be important to gain a better understanding of the role of this matter system, and its relation to the RNS fermions $\psi_\mu$.  In connecting down to 4d, reduction might seem a natural approach to understand the models of~\cite{Kunz:2020fwy,Kunz:2020rht}, and if so, their amplitudes. A further line is to find models that underpin the massive formul\ae\ in \cite{Albonico:2020mge}.

In the RNS model, the linearized field equation for external states came from requiring that the vertex operators were BRST invariant at the quantum level. The ambitwistor cohomology group $H^1(\P\A,\cO(2s-2))$ encodes generic space-time fields (not necessarily on-shell), so we expect quantum consistency of the current model to again impose field equations. In this paper, consistency was guaranteed by using vertex operators built from the Penrose of momentum eigenstate representatives on twistor space $Q$, which are necessarily on-shell. However, this was not obtained from a proper understanding of the BRST cohomology of the ambitwistor model, which is certainly desirable.

\medskip

The question arises as to whether 5d is special or whether one can construct analagous twistorial models in up to 10d as in the RNS case, or perhaps even 11d.  Such twistorial models have been investigated in higher dimensions, particularly in 6d as already discussed but also following the same logic as this paper, models for 10 and 11d were sketched  by two of the authors in  \cite{Geyer:2019ayz}. In 6, 10 and 11d, although the worldsheet theories were not set up completely systematically, a version sufficient to express the polarised scattering equations was introduced, together with supersymmetric amplitude formul\ae\ for gauge and gravity theories.  However, the worldsheet matter that would seem to give rise to the relevant formulae was not self-contained and seemed to require a doubling of the ambitwistor target space coordinates (and perhaps therefore also further constraints to compensate). Despite these hurdles, at some stage it may well become possible to complete  these models.
The 10 \& 11d models are rather large in that the little group in 10d is now SO(8) so that the analogue of the $a$ index ranges from 1 to 8, and 1 to 16 in 11d and hence we have many fields and constraints.   The models are closely related to models that had been introduced earlier as worldline models in \cite{Bars:2004dg,Bandos:2006nr,Uvarov:2015rxa} and proposed as ambitwistor strings in \cite{Bandos:2014lja}, see also  \cite{Bandos:2019zqp}, who proposes that the additional degrees of freedom correspond to those of M-theory.  

Different twistorial strategies are also possible in higher dimensions, in particular the worldline models of Berkovits    \cite{Berkovits:1990yc} is based on an impure twistor in 10 dimensions that nevertheless determines a null geodesic. This was elevated to an ambitwistor string in \cite{Berkovits:2019bbx,Reid-Edwards:2017goq}.  In 10d it is natural also to represent  null geodesics as intersections of pairs of $\alpha$-planes, and a model  based on two pure supertwistors  was proposed in  \cite{Sepulveda:2020kjc}.  
Given the parallels between conventional strings and ambitwistor strings, another area of investigation is to find twistorial representations of the standard superstring; work in this direction appears in \cite{Uvarov:2006ed,Uvarov:2007vs,Berkovits:2014aia,Berkovits:2015yra}.

\section*{Acknowledgements}
This project was initiated at the meeting at the Galileo Galilei Institute Florence at the workshop on String Theory from a Worldsheet Perspective 2019.  The work of YG is supported by the CUniverse research promotion project ``Toward World-Class Fundamental Physics'' of Chulalongkorn University (grant reference CUAASC).  LJM  is grateful to Mike Eastwood for useful correspondence on the 6d  Penrose transform,  and to the EPSRC for partial support   under grant EP/M018911/1 and  the STFC under  grant ST/T000864/1. DS is supported in part by STFC consolidated grant ST/P000681/1 and ST/T000694/1.  We are also grateful to the anonymous referee for pointing out glitches in the manuscript and raising  interesting points.

\appendix
\section{The polarized scattering equations and amplitude formulae }\label{polscatt}

We briefly review some of the material of \cite{ Geyer:2018xgb,Albonico:2020mge}
in order to motivate the form of the delta functions in 
\eqref{mom-estate} and to show how to obtain amplitude formulae. We first investigate how we can seek a spinor-helicity
factorization for  
\begin{equation}\label{P-def}
P_{AB}=\sum_{i=1}^n \frac{k_{iAB}}{(\sigma\sigma_i)}
\end{equation}over $\CP^1$; recall that if the scattering equations $k_i\cdot P(\sigma_i)=0$ hold $\forall i$, then $P(\sigma)\cdot P(\sigma)=0$ on $\CP^1$.  We can therefore hope to find $\lambda_{aA}(\sigma)$ and $\lambda^A_{\dot a}(\sigma)$  so that
\begin{equation}
P_{AB}=\frac{1}{2}\lambda_A^a\lambda_{Ba}=\frac{1}{2}\varepsilon_{ABCD}\lambda^C_{\dot a}\lambda^{D\dot a}\, .
\end{equation}  
The scattering equation $k_i\cdot P(\sigma_i)=0$ implies  $k_i\cdot P=\mathrm{det} (\kappa^a_{iA},\lambda^b_A)=0$.  This determinant  vanishes iff there exists non zero $(u^a_i,v^a_i)$ defined up to scale so that 
\begin{equation}
u_{ia}\lambda^a_A(\sigma_i) =v_{ia}\kappa_{iA}^a\, .
 \label{6dSE+}
\end{equation}
This is scale invariant in $u$ and $v$, so we can  normalize
\begin{equation}
\langle v\epsilon\rangle=1\, .\label{SE-norm}
\end{equation} 
The exponentials in the definition  of the vertex operators can be taken into the string path integral and leads to sources for the $\lambda_{aA}$ equations of motion in the form  
\begin{equation}
\dbar\lambda_{aA}= \sum_i u_{ia}\epsilon_{iA} \bar\delta((\sigma,\sigma_i))\, .
\end{equation}
Solving these gives rise  to an analogue of \eqref{P-def} (which arises in a similar way in the RNS ambitwistor string of \cite{Mason:2013sva}) for $\lambda_{Aa}(\sigma)$
\begin{equation}
\lambda_{aA}(\sigma)=\sum_{i=1}^n\frac{u_{ia}\epsilon_{iA}}{\sigma-\sigma_i}\, .\label{lambda-def}
\end{equation}
Together, \eqref{6dSE+}, \eqref{SE-norm} and \eqref{lambda-def} will constitute the \emph{polarized scattering equations}.  

These provide our $6d$ polarized version of the 4d polarized scattering equation of \cite{Geyer:2014fka} (referred to there as refined)  as equations on the $(\sigma_i, u_{ia}, v_{ia})$ that determine the $(u_{ia},v_{ia})$. 
More explicitly we can write
\begin{equation}
\sum_j \frac{\langle u_i u_j\rangle \epsilon_{jA}}{\sigma_{ij}}=\langle v_i \kappa_{iA} \rangle\, .\label{polscatt}
\end{equation}

It is shown in  \S3 of \cite{Albonico:2020mge} that generically there exists a unique solution to these equations for each solution $\sigma_i$ to the unpolarized scattering equation; the equations can be reduced to a linear system of the appropriate rank  for the $u_{ia}$s.

\paragraph{Measure:} After integrating out the exponentials in the vertex operators to obtain \eqref{lambda-def}, we are left with the integrations of the various parameters against the delta functions that   defines the measure based on the chiral 6D polarized scattering equations:
\begin{equation}
 \rd\mu_n^{\mathrm{pol}}= \frac{\prod_{i=1}^n  \delta\big(\langle v_i \varepsilon_i \rangle -1 \big)\;\delta^4\Big(\langle u_{i}\lambda_A(\sigma_i)\rangle -\langle v_{i}\kappa_{iA}\rangle\Big) \rd \sigma_i\,\rd^2 u_i\,\rd^2 v_i}{\mathrm{vol}\; \mathrm{SL}(2,\mathbb{C})_\sigma \times \mathrm{SL}(2,\mathbb{C})_+}\,. \label{measure6d}
\end{equation}
Here the  two copies of SL(2) are the Mobius transformations on $\sigma$ and the little group on the little $a$ index and the division by their volumes are   understood in the usual Faddeev-Popov sense.  In \S3 of \cite{Albonico:2020mge} it is shown that this measure is equal to the CHY measure. The correlator of the vertex operators reduces to the integral of the worldsheet correlators of the $w$'s against this measure. If one is only interested in amplitude formulae one can simply use the ingredients of the CHY formulae or the extra ingredients and alternatives presented in \cite{Geyer:2018xgb, Albonico:2020mge}.

\section{The third conformally invariant 6d ambitwistor string}\label{M-model}
As mentioned in \S\ref{sec:6d}, triality for $SO(8,\C)$, thought of as the conformal group of complexified 6d Minkowski space $\M$,  gives three representations of ambitwistor space as the space of complex projective lines, $\CP^1$s, in a 6-quadric. The main part of the text has been concerned with the case where the 6-quadric in question is $Q$, or $Q'$, the  twistor or primed twistor space where the quadrics are the projective  pure spinors  of $SO(8)$.  The third case of triality is when $\M$ itself is the quadric in $\P^7$ in the Cartan representation of the conformal group in which $SO(8)$ acts on the nonprojective space underlying $\P^7$ in the fundamental representation; this is usually known as the embedding formalism.
In this case the lines in $\M$ are the null geodesics themselves.  This particular model extends to give a conformally invariant  model for ambitwistor strings in arbitrary dimensions, although we will see that the anomaly calculation restricts us to 6d, at least for the bosonic theory.

In more detail, we extend the affine coordinates $x^{\mu}$, $\mu=0,\ldots ,d-1$ on $\M$ to $X^M\in \C^{d+2}$ equipped with an inner product by 
$$
 X^M =(x^{\mu},s,t)\, , \qquad X_1\cdot X_2= x^{\mu}_1x_{2\mu}+ s_1t_2+ t_1s_2\, .
$$
With this, we have the representation
\begin{equation}
\M=\left\{ X\in \C^8\ \middle|\ X\cdot X=0 \middle\}\right/ X\sim \alpha X\, , \qquad  \alpha\in \C\, .
\end{equation}
The conformal structure can be encoded in  the statement that two points, $X_1,X_2\in \M$ are null separated iff  $X_1\cdot X_2=0$.  In particular, points on the null geodesic connecting $X_1,X_2$ are simply in the linear span $X(u)=u^mX_m $ for some $u^m$, $m=1,2$;\footnote{Note that the $m$-index no longer has the interpretation as a little group index.} the null geodesic is represented by a null two-plane in $\C^{d+2}$.  Thus we arrive at a description of projective ambitwistor space as
\begin{equation}
\P\A= \left\{X_m\in \C^{d+2}\quad m=1,2\ \middle|\ X_m \cdot X_n=0\middle\}\right/\mathrm{GL}(2,\mathbb{C})\, .
\end{equation}
Here the $\mathrm{GL}(2,\mathbb{C})$ acts on the $m$-index, and for nonprojective ambitwistor space $\A$, we will only quotient by $\mathrm{SL}(2,\mathbb{C})$.

In order to compare to the standard affine representation of $\M$, we introduce the point at infinity 
$I^M=(0,0,1)$ so that points at null infinity $\scri$ are those  $X$ with $X\cdot I=0$.  With that finite points can be represented in terms of $x^{\mu}$ by $X=(x^{\mu}, 1, -x^2)$.  With this its easily checked that the symplectic potential $$
P\cdot dx= \varepsilon_{mn} X^m d X^n\, , \qquad \varepsilon_{mn}=\varepsilon_{[mn]}, \epsilon_{01}=1\,,
$$ 
in this presentation of $\P\A$ where we have set $X_1=(x^\mu,1,-x^2)$ and $X_2=(P^\mu , 0, -P\cdot x)$ with $P^2=0$, the intersection of the null geodesic with $\scri$. Thus, according to ambitwistor-lore, our model has worldsheet action
\begin{equation}
S=\int_\Sigma \varepsilon_{mn}X^m\bar D X^n \, , \qquad X_m \in \C^{2(d+2)}\otimes \sqrt{\Omega^{1,0}_\Sigma}\, ,
\end{equation}
where
$$
\bar D X_m =\dbar X_m  +A_m ^nX_n\,,  \qquad A_{mn}=A_{(mn)}\in \Omega^{0,1}_\Sigma \otimes sl_2\, .
$$
As before, the $A_{mn}$ are on the one hand gauge fields for the $\mathrm{SL}(2,\mathbb{C})$ quotient, and on the other  are  Lagrange multipliers that impose the constraints $X_m \cdot X_n=0$.

As before in equation \eqref{eqn:6d-gauge-anomaly}, this model has gauge anomalies in the quantum theory now of the form
\begin{equation}
\label{eqn:6d-spt-gauge-anomaly}
a_{\rm SL(2)} = \sum_{i} (-1)^F_i \,{\rm tr}_{R_i}(t^kt^k) = \frac{1}{2}\times (d+2)\,{\rm tr}_F(t^kt^k) - {\rm tr}_{\rm adj}(t^kt^k)\,,
\end{equation}
This calculation now shows that these cancel only when $d=6$.  This result is also achieved in the models of \cite{Adamo:2017zkm}; these are quite similar but with a different choice of worldsheet spins for the fields $X_1$ and $X_2$, i.e., one being a scalar and the other a $1$-form in the model of  \cite{Adamo:2017zkm}. 

\paragraph{Vertex operators.} Although this model has an identical structure to the twistor and models of either chirality   \eqref{eqn:ambitwistor-bosonic-action}, its relation to space-time is quite different. 
The distinction between these models becomes apparent when we come to the representation of vertex operators.  

In order to represent a momentum eigenstate, we need the point at infinity, $I=(0,0,1)$ as above, together with a  point $K=(k^\mu, 0,1)$ on its lightcone that must therefore satisfy $K\cdot I=0=K\cdot K=k\cdot k$. With this, an ordinary scalar momentum eigenstate is given by $e^{ik\cdot x}=\exp(iK\cdot X/I\cdot X)$.  For the standard integrated vertex operator of the bosonic RNS ambitwistor string \cite{Mason:2013sva}, 
\begin{equation}
V=\int w\,  \bar\delta(k\cdot P)e^{ik\cdot x}d\sigma,\label{normalVO}
\end{equation}
we observe that the delta function in enforcing the scattering equations in the form $k\cdot P=0$ guarantees  also that $k\cdot x$ is independent of the choice of $x$ on the null geodesic $x+u P$. This can be similarly enforced here by imposing the equations
$$
vX_m \cdot I + X_m \cdot K=0\, .
$$
This leads to the formulae
\begin{align}
V&=\int d\sigma\, dv \, w \,\bar  \delta^2(vX_m \cdot I + X_m \cdot K) \, \e^{iv} \label{eq:VO_X_delta}
\end{align}
and it is easily seen 
that by integrating out $v$ when $X_1=(x^\mu,1,-x^2)$ and $X_2=(P^\mu , 0, -P\cdot x)$ with $P^2=0$, we obtain the usual formula above. 

That $V$ in \eqref{eq:VO_X_delta} defines a consistent vertex operator  follows
because $I\cdot I=K\cdot K=I\cdot K=0$ implies that there are no normal ordering ambiguities.  The vertex operator is in the BRST cohomology because the current pulls out the argument of the delta-function,
\begin{equation*}
 X_1\cdot X_1(\sigma)\,\, \bar  \delta^2(\chi_I)(\sigma_m )\sim-\frac{1}{\sigma-\sigma_i}\underbrace{\big(X_1\cdot(v_m  I + K_m )\big)}_{=\chi_1}\, \bar  \delta(\chi_1)\, \bar  \delta'(\chi_2)\sim 0\,,
\end{equation*}
where we introduced the shorthand $\chi_I=X_I\cdot(v_m  I + K_m )$ for readability, and  higher order terms are absent due to $I\cdot I=K\cdot K=I\cdot K=0$. A similar argument ensures invariance under $X_2\cdot X_2$, whereas invariance under the off-diagonal current $X_1\cdot X_2$ follows from the antisymmetry of the $X_{1M}X_2^N$ OPE, together with $x\delta'(x)\simeq -\delta(x)$,
\begin{equation}
 X_1\cdot X_2(\sigma)\,\, \bar  \delta^2(\chi_I)(\sigma_i)\sim\frac{1}{\sigma-\sigma_i}\left(\chi_1\,\delta'(\chi_1)\delta(\chi_2)-\chi_2\,\delta(\chi_1)\delta'(\chi_2)\right)\sim 0\,.
\end{equation}

It is clear that for biadjoint scalar theory, $w$ should be given by a product of two independent current algebra generators. For the $(D^*F)^2$ theory and its double copy, the analogue of 
\label{DF2-Vop} here is to use $ E_{MN}X^M_mX^{Nm}$ in place of $E_{\cA\cB} Z^\cA_a Z^{\cB a}$ as these are the conformal group generators,  where $E_{MN}$ and $E_{\cA\cB}$ are identified as elements of the Lie algebra but expressed in the different  triality related  representations. 


For amplitude calculations, it is easy to see that the gauge fixings described above will reduce to the usual biadjoint scalar CHY formulae when $w$ is given by a product of two independent current algebra generators. 
When using $w= E_{MN}X^M_mX^N_n\varepsilon^{mn} J^at^a$ these give vertex operators that we expect to compute amplitudes for the $(D^*F)^2$ Lagrangian and its gravitational double copy.  We are not aware of attractive formulae for amplitudes at $n$-points for these latter theories.
It is an open question as to whether  representations can be found based on this model that differ  in an interesting way from the CHY formulae, except perhaps to make more manifest the action of conformal symmetry.

\section{Alternative worldsheet matter system}
\label{app:rho}
The worldsheet matter system $S_\rho$ of \eqref{eqn:rho-action} plays an important role in both physically interesting 5d models. Here, we present an alternative formulation of this model with fermions valued in the fundamental representation of the little group SL$(2,\mathbb{C})$,
\begin{equation}
 \tilde{S}_{\rho}=\int\frac{1}{2} \la\rho_\sA\,\dbar \rho_\sB\ra \,\Omega^{\sA\sB}+ \chi_{ab}\, \Omega^{\sA\sB}\lambda_\sA^a\rho_\sB^b+ \chi \,\Omega^{\sA\sB}\la\lambda_\sA\rho_\sB\ra\,.
\end{equation}
It should be intuitively clear that both actions are equivalent, since $\tilde{S}_{\rho}$ simply relabels the degrees of freedom and constraints. In particular, all critical models \eqref{eq:action_5d} could have been presented using $\tilde{S}_{\rho}$ instead of ${S}_{\rho}$.

 To spell this out in a bit more detail, let us BRST-quantize the resulting theories, introducing the  ghost systems $(\beta^{ab},\gamma_{ab})$ and $(\beta,\gamma)$ corresponding to $\chi_{ab}$ and $\chi$ respectively.  Since $\tilde{S}_{\rho}$ and ${S}_{\rho}$ have the same degrees of freedom with a given  worldsheet weight, the Virasoro anomaly counting agrees directly with \eqref{eq:Virasoro5d_sYM} $-$ \eqref{eq:Virasoro5d_sugra}. Moreover, the models using $\tilde{S}_\rho$ also have vanishing gauge anomalies, as can be seen e.g. from 
 \begin{equation}
 \mathfrak{a}_{\scalebox{0.7}{$\mathrm{SL}(2)$}}^{\scalebox{0.6}{{sYM}}}
 =\left( 4-\mathcal{N}\right)_\cZ\tr_{\scalebox{0.7}{F}}(t^kt^k)-2_\rho \,\tr_{\scalebox{0.7}{F}}(t^kt^k)+(1-1)\tr_{\scalebox{0.7}{adj}}(t^kt^k)=\frac{3}{2}\left(2-\mathcal{N}\right)
\end{equation}
in the case of {sYM}.
 
Following closely the discussion in \cref{sec:amplitudes}, we can again construct `picture (-1,-1)' super Yang-Mills vertex operators  from
\begin{equation}
 w_{\scalebox{0.6}{{sYM}}}^{(-1,-1)}\simeq\,\delta(\gamma)\,\delta\left(u_i^a u_i^b \gamma_{ab}\right)\;t_{\mathfrak{a}}j^{\mathfrak{a}}\,.
\end{equation}
As before, these are in the BRST cohomology due to the invariance of the bare vertex operators \eqref{super-mom-estate} under $\la u\lambda_\sA\ra\rho^{\sA b}$, and so carry only two delta-functions. The picture numbers $(-1,-1)$ refer to the $\beta\gamma$ and $\beta^{ab}\gamma_{ab}$ systems respectively. Picture changing operators can be derived following the discussion in \cref{sec:amplitudes}, 
\begin{equation}
 \Upsilon(z_l)=\delta(\beta)\,\la\lambda_\sA\rho_\sB\ra\Omega^{\sA\sB}\,,\qquad \hat \Upsilon(z_l)=\delta\big(\hat u_a\hat u_b \beta^{ab}\big)\, \la \hat u \lambda_\sA\ra\la \hat u \rho_\sB\ra\,\Omega^{\sA\sB}\,,
\end{equation}
with $n-1$ insertions of $\Upsilon$ and $n-3$ insertions of $\hat \Upsilon$ in an $n$-point correlator, matching the zero mode counting $n_\gamma=n_{\gamma_{ab}}=1$. Using the localization condition \eqref{eq:sol_lambda} for $\lambda$ this leads to the following integrated vertex operators,
\begin{equation}
 w_{\scalebox{0.6}{sYM}}^{(0,-1)}\simeq\delta\left(u_i^a u_i^b \gamma_{ab}\right)\, \la u_i \rho_\sA\ra \epsilon_i^\sA\;t_{\mathfrak{a}}j^{\mathfrak{a}} \,,\qquad
 w_{\scalebox{0.6}{sYM}}^{(0,0)}\simeq\,
 \, \left(e_i\cdot P(\sigma_i) +\la \rho_\sA\rho_\sB\ra \epsilon_i^\sA\epsilon_i^\sB\right)\;t_{\mathfrak{a}}j^{\mathfrak{a}}
 \,,
\end{equation}
where correlators only give non-trivial results of the picture numbers add up to $(-1,-3)$. At this point, it is can easily be checked that correlators agree with the $S_\rho$ system: the ghost correlators contribute
\begin{equation}
 \left\langle \delta(\gamma)\, \delta\left(u_1^a u_1^b \gamma_{ab}\right)\delta\left(u_2^a u_2^b \gamma_{ab}\right)\delta\left(u_3^a u_3^b \gamma_{ab}\right)\right\rangle = \frac{1}{\la u_1 u_2\ra\,\la u_2 u_3\ra\,\la u_3 u_1\ra}\,,
\end{equation}
which combines with the $\rho$-correlator,
\begin{equation}
 \left\langle  \la u_2 \rho_\sA\ra \epsilon_2^\sA\,
 \la u_3 \rho_\sB\ra \epsilon_3^\sB\,
 \prod_{i=4}^n \left(e_i\cdot P(\sigma_i) +\la \rho_{\sA_i}\rho_{\sB_i}\ra \epsilon_i^{\sA_i}\epsilon_i^{\sB_i}\right)
 \right\rangle = \la u_2 u_3\ra\; \mathrm{det} H^{[12]}_{[13]}\,.
\end{equation}
to give the familiar reduced determinant;
\begin{equation}
 \mathcal{I}_n=\left\langle w_1^{(-1,-1)} w_2^{(0,-1)}w_3^{(0,-1)}\prod_{i=4}^n w_i^{(0,0)} \right\rangle=\det{}'H\,.
\end{equation}

\bibliography{twistor-bib}

\providecommand{\href}[2]{#2}\begingroup\raggedright\begin{thebibliography}{10}

\bibitem{Mason:2013sva}
L.~Mason and D.~Skinner, \emph{{Ambitwistor strings and the scattering
  equations}}, \href{http://dx.doi.org/10.1007/JHEP07(2014)048}{\emph{JHEP}
  {\bfseries 1407} (2014) 048},
  [\href{https://arxiv.org/abs/1311.2564}{{\ttfamily 1311.2564}}].

\bibitem{Witten:2003nn}
E.~Witten, \emph{{Perturbative gauge theory as a string theory in twistor
  space}},
  \href{http://dx.doi.org/10.1007/s00220-004-1187-3}{\emph{Commun.Math.Phys.}
  {\bfseries 252} (2004) 189--258},
  [\href{https://arxiv.org/abs/hep-th/0312171}{{\ttfamily hep-th/0312171}}].

\bibitem{Berkovits:2004hg}
N.~Berkovits, \emph{{An Alternative string theory in twistor space for N=4
  superYang-Mills}},
  \href{http://dx.doi.org/10.1103/PhysRevLett.93.011601}{\emph{Phys.Rev.Lett.}
  {\bfseries 93} (2004) 011601},
  [\href{https://arxiv.org/abs/hep-th/0402045}{{\ttfamily hep-th/0402045}}].

\bibitem{Berkovits:2013xba}
N.~Berkovits, \emph{{Infinite Tension Limit of the Pure Spinor Superstring}},
  \href{http://dx.doi.org/10.1007/JHEP03(2014)017}{\emph{JHEP} {\bfseries 1403}
  (2014) 017}, [\href{https://arxiv.org/abs/1311.4156}{{\ttfamily 1311.4156}}].

\bibitem{Geyer:2014fka}
Y.~Geyer, A.~E. Lipstein and L.~J. Mason, \emph{{Ambitwistor Strings in Four
  Dimensions}},
  \href{http://dx.doi.org/10.1103/PhysRevLett.113.081602}{\emph{Phys. Rev.
  Lett.} {\bfseries 113} (2014) 081602},
  [\href{https://arxiv.org/abs/1404.6219}{{\ttfamily 1404.6219}}].

\bibitem{Casali:2015vta}
E.~Casali, Y.~Geyer, L.~Mason, R.~Monteiro and K.~A. Roehrig, \emph{{New
  Ambitwistor String Theories}},
  \href{http://dx.doi.org/10.1007/JHEP11(2015)038}{\emph{JHEP} {\bfseries 11}
  (2015) 038}, [\href{https://arxiv.org/abs/1506.08771}{{\ttfamily
  1506.08771}}].

\bibitem{Cachazo:2013iea}
F.~Cachazo, S.~He and E.~Y. Yuan, \emph{{Scattering of Massless Particles:
  Scalars, Gluons and Gravitons}},
  \href{http://dx.doi.org/10.1007/JHEP07(2014)033}{\emph{JHEP} {\bfseries 1407}
  (2014) 033}, [\href{https://arxiv.org/abs/1309.0885}{{\ttfamily 1309.0885}}].

\bibitem{Adamo:2013tsa}
T.~Adamo, E.~Casali and D.~Skinner, \emph{{Ambitwistor strings and the
  scattering equations at one loop}},
  \href{http://dx.doi.org/10.1007/JHEP04(2014)104}{\emph{JHEP} {\bfseries 1404}
  (2014) 104}, [\href{https://arxiv.org/abs/1312.3828}{{\ttfamily 1312.3828}}].

\bibitem{Roiban:2004yf}
R.~Roiban, M.~Spradlin and A.~Volovich, \emph{{On the tree level S matrix of
  Yang-Mills theory}},
  \href{http://dx.doi.org/10.1103/PhysRevD.70.026009}{\emph{Phys.Rev.}
  {\bfseries D70} (2004) 026009},
  [\href{https://arxiv.org/abs/hep-th/0403190}{{\ttfamily hep-th/0403190}}].

\bibitem{Skinner:2013xp}
D.~Skinner, \emph{{Twistor Strings for N=8 Supergravity}},
  \href{https://arxiv.org/abs/1301.0868}{{\ttfamily 1301.0868}}.

\bibitem{Boothby:1961}
W.~M. Boothby, \emph{Homogeneous complex contact manifolds}, {\emph{Proc. Symp.
  Pure Math. III (Differential Geometry)} (1961) 144-- 154}.

\bibitem{Baston:1989vh}
R.~Baston and M.~G. Eastwood, \emph{{The Penrose transform: Its interaction
  with representation theory}}.
\newblock OUP, 1989.

\bibitem{Penrose:1986ca}
R.~Penrose and W.~Rindler, \emph{{Spinors and Space-Time: Spinor And Twistor
  Methods In Space-time Geometry}}, vol.~2.
\newblock Cambridge University Press, 1986.

\bibitem{Hughston:1988nz}
L.~Hughston and L.~Mason, \emph{{A Generalized Kerr-Robinson Theorem}},
  \href{http://dx.doi.org/10.1088/0264-9381/5/2/007}{\emph{Class. Quant. Grav.}
  {\bfseries 5} (1988) 275--285}.

\bibitem{Harnad:1987xq}
J.~P. Harnad and S.~Shnider, \emph{{Isotropic geometry, Twistors and
  Supertwistors. 1. The generalized Klein correspondence and Spinor Flags}},
  \href{http://dx.doi.org/10.1063/1.529538}{\emph{J. Math. Phys.} {\bfseries
  33} (1992) 3197--3208}.

\bibitem{Harnad:1995zy}
J.~P. Harnad and S.~Shnider, \emph{{Isotropic geometry and twistors in higher
  dimensions. 2: Odd dimensions, reality conditions, and twistor superspaces}},
  \href{http://dx.doi.org/10.1063/1.531096}{\emph{J. Math. Phys.} {\bfseries
  36} (1995) 1945--1970}.

\bibitem{Geyer:2018xgb}
Y.~Geyer and L.~Mason, \emph{{Polarized Scattering Equations for 6D
  Superamplitudes}},
  \href{http://dx.doi.org/10.1103/PhysRevLett.122.101601}{\emph{Phys. Rev.
  Lett.} {\bfseries 122} (2019) 101601},
  [\href{https://arxiv.org/abs/1812.05548}{{\ttfamily 1812.05548}}].

\bibitem{Geyer:2019ayz}
Y.~Geyer and L.~Mason, \emph{{Supersymmetric S-matrices from the worldsheet in
  10 \textbackslash{}\& 11d}},
  \href{http://dx.doi.org/10.1016/j.physletb.2020.135361}{\emph{Phys. Lett. B}
  {\bfseries 804} (2020) 135361},
  [\href{https://arxiv.org/abs/1901.00134}{{\ttfamily 1901.00134}}].

\bibitem{Johansson:2017srf}
H.~Johansson and J.~Nohle, \emph{{Conformal Gravity from Gauge Theory}},
  \href{https://arxiv.org/abs/1707.02965}{{\ttfamily 1707.02965}}.

\bibitem{Adamo:2017zkm}
T.~Adamo, R.~Monteiro and M.~F. Paulos, \emph{{Space-time CFTs from the Riemann
  sphere}}, \href{http://dx.doi.org/10.1007/JHEP08(2017)067}{\emph{JHEP}
  {\bfseries 08} (2017) 067},
  [\href{https://arxiv.org/abs/1703.04589}{{\ttfamily 1703.04589}}].

\bibitem{Mason:2011nw}
L.~J. Mason, R.~A. Reid-Edwards and A.~Taghavi-Chabert, \emph{{Conformal Field
  Theories in Six-Dimensional Twistor Space}},
  \href{http://dx.doi.org/10.1016/j.geomphys.2012.08.001}{\emph{J. Geom. Phys.}
  {\bfseries 62} (2012) 2353--2375},
  [\href{https://arxiv.org/abs/1111.2585}{{\ttfamily 1111.2585}}].

\bibitem{Cachazo:2013hca}
F.~Cachazo, S.~He and E.~Y. Yuan, \emph{{Scattering of Massless Particles in
  Arbitrary Dimensions}},
  \href{http://dx.doi.org/10.1103/PhysRevLett.113.171601}{\emph{Phys.Rev.Lett.}
  {\bfseries 113} (2014) 171601},
  [\href{https://arxiv.org/abs/1307.2199}{{\ttfamily 1307.2199}}].

\bibitem{Albonico:2020mge}
G.~Albonico, Y.~Geyer and L.~Mason, \emph{{Recursion and worldsheet formulae
  for 6d superamplitudes}},
  \href{http://dx.doi.org/10.1007/JHEP08(2020)066}{\emph{JHEP} {\bfseries 08}
  (2020) 066}, [\href{https://arxiv.org/abs/2001.05928}{{\ttfamily
  2001.05928}}].

\bibitem{Chern:2009nt}
T.~Chern, \emph{{Superconformal Field Theory In Six Dimensions And
  Supertwistor}},  \href{https://arxiv.org/abs/0906.0657}{{\ttfamily
  0906.0657}}.

\bibitem{Mason:2012va}
L.~J. Mason and R.~A. Reid-Edwards, \emph{{The supersymmetric Penrose transform
  in six dimensions}},  \href{https://arxiv.org/abs/1212.6173}{{\ttfamily
  1212.6173}}.

\bibitem{Azevedo:2017lkz}
T.~Azevedo and O.~T. Engelund, \emph{{Ambitwistor formulations of R$^{2}$
  gravity and (DF)$^{2}$ gauge theories}},
  \href{http://dx.doi.org/10.1007/JHEP11(2017)052}{\emph{JHEP} {\bfseries 11}
  (2017) 052}, [\href{https://arxiv.org/abs/1707.02192}{{\ttfamily
  1707.02192}}].

\bibitem{Berkovits:2018jvm}
N.~Berkovits and M.~Lize, \emph{{Field theory actions for ambitwistor string
  and superstring}},
  \href{http://dx.doi.org/10.1007/JHEP09(2018)097}{\emph{JHEP} {\bfseries 09}
  (2018) 097}, [\href{https://arxiv.org/abs/1807.07661}{{\ttfamily
  1807.07661}}].

\bibitem{Azevedo:2019zbn}
T.~Azevedo, R.~L. Jusinskas and M.~Lize, \emph{{Bosonic sectorized strings and
  the $(DF)^{2}$ theory}},
  \href{http://dx.doi.org/10.1007/JHEP01(2020)082}{\emph{JHEP} {\bfseries 01}
  (2020) 082}, [\href{https://arxiv.org/abs/1908.11371}{{\ttfamily
  1908.11371}}].

\bibitem{Branson:2004}
T.~Branson and A.~R. Gover, \emph{Conformally invariant operators, differential
  forms, cohomology and a generalisation of q-curvature},
  \href{http://dx.doi.org/10.1080/03605300500299943}{\emph{Communications in
  Partial Differential Equations} {\bfseries 30} (2005) 1611--1669},
  [\href{https://arxiv.org/abs/https://doi.org/10.1080/03605300500299943}{{\ttfamily
  https://doi.org/10.1080/03605300500299943}}].

\bibitem{Cheung:2009dc}
C.~Cheung and D.~O'Connell, \emph{{Amplitudes and Spinor-Helicity in Six
  Dimensions}},
  \href{http://dx.doi.org/10.1088/1126-6708/2009/07/075}{\emph{JHEP} {\bfseries
  07} (2009) 075}, [\href{https://arxiv.org/abs/0902.0981}{{\ttfamily
  0902.0981}}].

\bibitem{Dennen:2009vk}
T.~Dennen, Y.-t. Huang and W.~Siegel, \emph{{Supertwistor space for 6D maximal
  super Yang-Mills}},
  \href{http://dx.doi.org/10.1007/JHEP04(2010)127}{\emph{JHEP} {\bfseries 04}
  (2010) 127}, [\href{https://arxiv.org/abs/0910.2688}{{\ttfamily 0910.2688}}].

\bibitem{Huang:2010rn}
Y.-t. Huang and A.~E. Lipstein, \emph{{Amplitudes of 3D and 6D Maximal
  Superconformal Theories in Supertwistor Space}},
  \href{http://dx.doi.org/10.1007/JHEP10(2010)007}{\emph{JHEP} {\bfseries 10}
  (2010) 007}, [\href{https://arxiv.org/abs/1004.4735}{{\ttfamily 1004.4735}}].

\bibitem{Cachazo:2018hqa}
F.~Cachazo, A.~Guevara, M.~Heydeman, S.~Mizera, J.~H. Schwarz and C.~Wen,
  \emph{{The S Matrix of 6D Super Yang-Mills and Maximal Supergravity from
  Rational Maps}}, \href{http://dx.doi.org/10.1007/JHEP09(2018)125}{\emph{JHEP}
  {\bfseries 09} (2018) 125},
  [\href{https://arxiv.org/abs/1805.11111}{{\ttfamily 1805.11111}}].

\bibitem{Heydeman:2017yww}
M.~Heydeman, J.~H. Schwarz and C.~Wen, \emph{{M5-Brane and D-Brane Scattering
  Amplitudes}}, \href{http://dx.doi.org/10.1007/JHEP12(2017)003}{\emph{JHEP}
  {\bfseries 12} (2017) 003},
  [\href{https://arxiv.org/abs/1710.02170}{{\ttfamily 1710.02170}}].

\bibitem{Witten:2012bh}
E.~Witten, \emph{{Superstring Perturbation Theory Revisited}},
  \href{https://arxiv.org/abs/1209.5461}{{\ttfamily 1209.5461}}.

\bibitem{Kunz:2020fwy}
C.~Kunz, \emph{{Four Dimensional Anomaly-Free Twistor String}},
  \href{https://arxiv.org/abs/2004.04842}{{\ttfamily 2004.04842}}.

\bibitem{Kunz:2020rht}
C.~Kunz, \emph{{Spectrum, Scattering Amplitudes, and String Field Theory of 4
  Dimensional Twistor String}},
  \href{https://arxiv.org/abs/2010.02284}{{\ttfamily 2010.02284}}.

\bibitem{Bars:2004dg}
I.~Bars, \emph{{Twistor superstring in 2T-physics}},
  \href{http://dx.doi.org/10.1103/PhysRevD.70.104022}{\emph{Phys. Rev. D}
  {\bfseries 70} (2004) 104022},
  [\href{https://arxiv.org/abs/hep-th/0407239}{{\ttfamily hep-th/0407239}}].

\bibitem{Bandos:2006nr}
I.~A. Bandos, J.~A. de~Azcarraga and D.~P. Sorokin, \emph{{On D=11
  supertwistors, superparticle quantization and a hidden SO(16) symmetry of
  supergravity}},  in \emph{{Proceedings, 22nd Max Born Symposium on Quantum,
  Super and Twistors: A Conference in Honor of Jerzy Lukierski on His 70th
  Birthday: Wroclaw, Poland, September 27-29, 2006}}, 2006,
  \href{https://arxiv.org/abs/hep-th/0612252}{{\ttfamily hep-th/0612252}}.

\bibitem{Uvarov:2015rxa}
D.~Uvarov, \emph{{Spinor description of D = 5 massless low-spin gauge fields}},
  \href{http://dx.doi.org/10.1088/0264-9381/33/13/135010}{\emph{Class. Quant.
  Grav.} {\bfseries 33} (2016) 135010},
  [\href{https://arxiv.org/abs/1506.01881}{{\ttfamily 1506.01881}}].

\bibitem{Bandos:2014lja}
I.~Bandos, \emph{{Twistor/ambitwistor strings and null-superstrings in
  spacetime of D=4, 10 and 11 dimensions}},
  \href{http://dx.doi.org/10.1007/JHEP09(2014)086}{\emph{JHEP} {\bfseries 09}
  (2014) 086}, [\href{https://arxiv.org/abs/1404.1299}{{\ttfamily 1404.1299}}].

\bibitem{Bandos:2019zqp}
I.~Bandos, \emph{{On polarized scattering equations for superamplitudes of 11D
  supergravity and ambitwistor superstring}},
  \href{http://dx.doi.org/10.1007/JHEP11(2019)087}{\emph{JHEP} {\bfseries 11}
  (2019) 087}, [\href{https://arxiv.org/abs/1908.07482}{{\ttfamily
  1908.07482}}].

\bibitem{Berkovits:1990yc}
N.~Berkovits, \emph{{A Supertwistor Description of the Massless Superparticle
  in Ten-dimensional Superspace}},
  \href{http://dx.doi.org/10.1016/0370-2693(90)91047-F}{\emph{Phys. Lett. B}
  {\bfseries 247} (1990) 45--49}.

\bibitem{Berkovits:2019bbx}
N.~Berkovits, M.~Guillen and L.~Mason, \emph{{Supertwistor description of
  ambitwistor strings}},
  \href{http://dx.doi.org/10.1007/JHEP01(2020)020}{\emph{JHEP} {\bfseries 01}
  (2020) 020}, [\href{https://arxiv.org/abs/1908.06899}{{\ttfamily
  1908.06899}}].

\bibitem{Reid-Edwards:2017goq}
R.~Reid-Edwards and D.~Riccombeni, \emph{{A Superstring Field Theory for
  Supergravity}}, \href{http://dx.doi.org/10.1007/JHEP09(2017)103}{\emph{JHEP}
  {\bfseries 09} (2017) 103},
  [\href{https://arxiv.org/abs/1701.05495}{{\ttfamily 1701.05495}}].

\bibitem{Sepulveda:2020kjc}
D.~Garc\'\i{}a~Sep\'ulveda and M.~Guillen, \emph{{A pure spinor twistor
  description of the $D = 10$ superparticle}},
  \href{http://dx.doi.org/10.1007/JHEP08(2020)130}{\emph{JHEP} {\bfseries 08}
  (2020) 130}, [\href{https://arxiv.org/abs/2006.06023}{{\ttfamily
  2006.06023}}].

\bibitem{Uvarov:2006ed}
D.~Uvarov, \emph{{(Super)twistors and (super)strings}},
  \href{http://dx.doi.org/10.1088/0264-9381/23/7/029}{\emph{Class. Quant.
  Grav.} {\bfseries 23} (2006) 2711--2726},
  [\href{https://arxiv.org/abs/hep-th/0601149}{{\ttfamily hep-th/0601149}}].

\bibitem{Uvarov:2007vs}
D.~Uvarov, \emph{{Supertwistor formulation for higher dimensional
  superstrings}},
  \href{http://dx.doi.org/10.1088/0264-9381/24/22/004}{\emph{Class. Quant.
  Grav.} {\bfseries 24} (2007) 5383--5400},
  [\href{https://arxiv.org/abs/hep-th/0703051}{{\ttfamily hep-th/0703051}}].

\bibitem{Berkovits:2014aia}
N.~Berkovits, \emph{{Twistor Origin of the Superstring}},
  \href{http://dx.doi.org/10.1007/JHEP03(2015)122}{\emph{JHEP} {\bfseries 03}
  (2015) 122}, [\href{https://arxiv.org/abs/1409.2510}{{\ttfamily 1409.2510}}].

\bibitem{Berkovits:2015yra}
N.~Berkovits, \emph{{Origin of the Pure Spinor and Green-Schwarz Formalisms}},
  \href{http://dx.doi.org/10.1007/JHEP07(2015)091}{\emph{JHEP} {\bfseries 07}
  (2015) 091}, [\href{https://arxiv.org/abs/1503.03080}{{\ttfamily
  1503.03080}}].

\end{thebibliography}\endgroup
\bibliographystyle{JHEP}

\end{document}